\pdfoutput=1
\documentclass[reprint,aps,prd,superscriptaddress,showpacs,preprintnumbers,amsmath,amssymb,floatfix]{revtex4-1} 

\usepackage[mathlines]{lineno}
\usepackage{graphicx,subfigure}
\usepackage{epstopdf}
\usepackage{dcolumn}
\usepackage{bm}
\usepackage{color}
\usepackage{upgreek}
\usepackage{afterpage}
\usepackage{tikz, pgfplots}
\usepackage{mhchem} 
\usepackage[colorlinks=true]{hyperref}

\def\ne#1#2{#1\,$\pm$\,#2}
\def\nes#1#2#3{#1\,$\pm$\,#2$_\mathrm{stat}$\,$\pm$\,#3$_\mathrm{sys}$}
\def\nae#1#2#3{#1\,(+#2\,-#3)$_{\mathrm{stat}}$}

\newcommand{\eq}{Eq.}
\newcommand{\peq}{Eqs.}
\newcommand{\fig}{Fig.}

\newcommand{\tab}{Table}

\newcommand{\sect}{Sec.}
\newcommand{\psect}{Secs.}

\newcommand{\gev}{GeV/$c^2$}
\newcommand{\keVr}{keV} 
\newcommand{\keVee}{keV$_{\text{ee}}$} 
\newcommand{\perkgd}{/kg/day} 
\newcommand{\kgd}{kg\,days} 
\newcommand{\mus}{\ensuremath{\mu{}s}} 

\newcommand{\fivedchi}{5d-\ensuremath{\chi^2}}
\newcommand{\fivedchiex}{extended 5d-\ensuremath{\chi^2}}
\newcommand{\cdmstwo}{CDMS~II}

\newcommand{\se}{surface-event}
\newcommand{\seuh}{surface event}
\newcommand{\er}{ER}
\newcommand{\nr}{NR}
\begin{document}

\preprint{CDMS \today}

\title{Improved WIMP-search reach of the \cdmstwo{} germanium data
}

\affiliation{Division of Physics, Mathematics, \& Astronomy, California Institute of Technology, Pasadena, California 91125, USA} 
\affiliation{Institute for Particle Physics Phenomenology, Department of Physics, Durham University, Durham, United Kingdom} 
\affiliation{Fermi National Accelerator Laboratory, Batavia, Illinois 60510, USA} 
\affiliation{Lawrence Berkeley National Laboratory, Berkeley, California 94720, USA} 
\affiliation{Department of Physics, Massachusetts Institute of Technology, Cambridge, Massachusetts 02139, USA} 
\affiliation{Pacific Northwest National Laboratory, Richland, Washington 99352, USA} 
\affiliation{Department of Physics, Queen's University, Kingston Ontario K7L 3N6, Canada} 
\affiliation{Department of Physics, Santa Clara University, Santa Clara, California 95053, USA} 
\affiliation{SLAC National Accelerator Laboratory/Kavli Institute for Particle Astrophysics and Cosmology, Menlo Park, California 94025, USA} 
\affiliation{Department of Physics, South Dakota School of Mines and Technology, Rapid City, South Dakota 57701, USA} 
\affiliation{Department of Physics, Southern Methodist University, Dallas, Texas 75275, USA}
\affiliation{Department of Physics, Stanford University, Stanford, California 94305, USA} 
\affiliation{Department of Physics, Syracuse University, Syracuse, New York 13244, USA} 
\affiliation{Department of Physics and Astronomy, and the Mitchell Institute for Fundamental Physics and Astronomy, Texas A\&M University, College Station, Texas 77843, USA} 
\affiliation{Departamento de F\'{\i}sica Te\'orica and Instituto de F\'{\i}sica Te\'orica UAM/CSIC, Universidad Aut\'onoma de Madrid, 28049 Madrid, Spain} 
\affiliation{Department of Physics \& Astronomy, University of British Columbia, Vancouver,  British Columbia  V6T 1Z1, Canada} 
\affiliation{Department of Physics, University of California, Berkeley, California 94720, USA} 
\affiliation{Department of Physics, University of California, Santa Barbara, California 93106, USA} 
\affiliation{Department of Physics, University of Colorado Denver, Denver, Colorado 80217, USA} 
\affiliation{Department of Physics, University of Evansville, Evansville, Indiana 47722, USA} 
\affiliation{Department of Physics, University of Florida, Gainesville, Florida 32611, USA} 
\affiliation{Department of Physics, University of Illinois at Urbana-Champaign, Urbana, Illinois 61801, USA} 
\affiliation{School of Physics \& Astronomy, University of Minnesota, Minneapolis, Minnesota 55455, USA} 
\affiliation{Department of Physics, University of South Dakota, Vermillion, South Dakota 57069, USA} 

\author{R.~Agnese} \affiliation{Department of Physics, University of Florida, Gainesville, FL 32611, USA} 
\author{A.J.~Anderson} \affiliation{Department of Physics, Massachusetts Institute of Technology, Cambridge, MA 02139, USA} 
\author{M.~Asai} \affiliation{SLAC National Accelerator Laboratory/Kavli Institute for Particle Astrophysics and Cosmology, Menlo Park 94025, CA} 
\author{D.~Balakishiyeva} \affiliation{Department of Physics, University of Florida, Gainesville, FL 32611, USA} 
\author{D.~Barker} \affiliation{School of Physics \& Astronomy, University of Minnesota, Minneapolis, MN 55455, USA} 
\author{R.~Basu~Thakur~} \affiliation{Fermi National Accelerator Laboratory, Batavia, IL 60510, USA}\affiliation{Department of Physics, University of Illinois at Urbana-Champaign, Urbana, IL 61801, USA} 
\author{D.A.~Bauer} \affiliation{Fermi National Accelerator Laboratory, Batavia, IL 60510, USA} 
\author{J.~Billard} \affiliation{Department of Physics, Massachusetts Institute of Technology, Cambridge, MA 02139, USA} 
\author{A.~Borgland} \affiliation{SLAC National Accelerator Laboratory/Kavli Institute for Particle Astrophysics and Cosmology, Menlo Park 94025, CA} 
\author{M.A.~Bowles} \affiliation{Department of Physics, Syracuse University, Syracuse, NY 13244, USA} 
\author{D.~Brandt} \affiliation{SLAC National Accelerator Laboratory/Kavli Institute for Particle Astrophysics and Cosmology, Menlo Park 94025, CA} 
\author{P.L.~Brink} \affiliation{SLAC National Accelerator Laboratory/Kavli Institute for Particle Astrophysics and Cosmology, Menlo Park 94025, CA} 
\author{R.~Bunker} \affiliation{Department of Physics, South Dakota School of Mines and Technology, Rapid City, SD 57701, USA} 
\author{B.~Cabrera} \affiliation{Department of Physics, Stanford University, Stanford, CA 94305, USA} 
\author{D.O.~Caldwell} \affiliation{Department of Physics, University of California, Santa Barbara, CA 93106, USA} 
\author{R.~Calkins} \affiliation{Department of Physics, Southern Methodist University, Dallas, TX 75275, USA} 
\author{D.G.~Cerde\~no} \affiliation{Institute for Particle Physics Phenomenology, Department of Physics, Durham University, Durham, UK} 
\author{H.~Chagani} \affiliation{School of Physics \& Astronomy, University of Minnesota, Minneapolis, MN 55455, USA} 
\author{Y.~Chen} \affiliation{Department of Physics, Syracuse University, Syracuse, NY 13244, USA} 
\author{J.~Cooley} \affiliation{Department of Physics, Southern Methodist University, Dallas, TX 75275, USA} 
\author{B.~Cornell} \affiliation{Division of Physics, Mathematics, \& Astronomy, California Institute of Technology, Pasadena, CA 91125, USA} 
\author{C.H.~Crewdson} \affiliation{Department of Physics, Queen's University, Kingston ON, Canada K7L 3N6} 
\author{P.~Cushman} \affiliation{School of Physics \& Astronomy, University of Minnesota, Minneapolis, MN 55455, USA} 
\author{M.~Daal} \affiliation{Department of Physics, University of California, Berkeley, CA 94720, USA} 
\author{P.C.F.~Di~Stefano} \affiliation{Department of Physics, Queen's University, Kingston ON, Canada K7L 3N6} 
\author{T.~Doughty} \affiliation{Department of Physics, University of California, Berkeley, CA 94720, USA} 
\author{L.~Esteban} \affiliation{Departamento de F\'{\i}sica Te\'orica and Instituto de F\'{\i}sica Te\'orica UAM/CSIC, Universidad Aut\'onoma de Madrid, 28049 Madrid, Spain} 
\author{S.~Fallows} \affiliation{School of Physics \& Astronomy, University of Minnesota, Minneapolis, MN 55455, USA} 
\author{E.~Figueroa-Feliciano} \affiliation{Department of Physics, Massachusetts Institute of Technology, Cambridge, MA 02139, USA} 
\author{G.L.~Godfrey} \affiliation{SLAC National Accelerator Laboratory/Kavli Institute for Particle Astrophysics and Cosmology, Menlo Park 94025, CA} 
\author{S.R.~Golwala} \affiliation{Division of Physics, Mathematics, \& Astronomy, California Institute of Technology, Pasadena, CA 91125, USA} 
\author{J.~Hall} \affiliation{Pacific Northwest National Laboratory, Richland, WA 99352, USA} 
\author{H.R.~Harris} \affiliation{Department of Physics and Astronomy, and the Mitchell Institute for Fundamental Physics and Astronomy, Texas A\&M University, College Station, TX 77843, USA} 
\author{S.A.~Hertel} \affiliation{Department of Physics, Massachusetts Institute of Technology, Cambridge, MA 02139, USA} 
\author{T.~Hofer} \affiliation{School of Physics \& Astronomy, University of Minnesota, Minneapolis, MN 55455, USA} 
\author{D.~Holmgren} \affiliation{Fermi National Accelerator Laboratory, Batavia, IL 60510, USA} 
\author{L.~Hsu} \affiliation{Fermi National Accelerator Laboratory, Batavia, IL 60510, USA} 
\author{M.E.~Huber} \affiliation{Department of Physics, University of Colorado Denver, Denver, CO 80217, USA} 
\author{D.~Jardin} \affiliation{Department of Physics, Southern Methodist University, Dallas, TX 75275, USA}
\author{A.~Jastram} \affiliation{Department of Physics and Astronomy, and the Mitchell Institute for Fundamental Physics and Astronomy, Texas A\&M University, College Station, TX 77843, USA} 
\author{O.~Kamaev} \affiliation{Department of Physics, Queen's University, Kingston ON, Canada K7L 3N6} 
\author{B.~Kara} \affiliation{Department of Physics, Southern Methodist University, Dallas, TX 75275, USA} 
\author{M.H.~Kelsey} \affiliation{SLAC National Accelerator Laboratory/Kavli Institute for Particle Astrophysics and Cosmology, Menlo Park 94025, CA} 
\author{A.~Kennedy} \affiliation{School of Physics \& Astronomy, University of Minnesota, Minneapolis, MN 55455, USA} 
\author{M.~Kiveni} \affiliation{Department of Physics, Syracuse University, Syracuse, NY 13244, USA} 
\author{K.~Koch} \affiliation{School of Physics \& Astronomy, University of Minnesota, Minneapolis, MN 55455, USA} 
\author{A.~Leder} \affiliation{Department of Physics, Massachusetts Institute of Technology, Cambridge, MA 02139, USA} 
\author{B.~Loer} \affiliation{Fermi National Accelerator Laboratory, Batavia, IL 60510, USA} 
\author{E.~Lopez~Asamar} \affiliation{Departamento de F\'{\i}sica Te\'orica and Instituto de F\'{\i}sica Te\'orica UAM/CSIC, Universidad Aut\'onoma de Madrid, 28049 Madrid, Spain} 
\author{P.~Lukens} \affiliation{Fermi National Accelerator Laboratory, Batavia, IL 60510, USA} 
\author{R.~Mahapatra} \affiliation{Department of Physics and Astronomy, and the Mitchell Institute for Fundamental Physics and Astronomy, Texas A\&M University, College Station, TX 77843, USA} 
\author{V.~Mandic} \affiliation{School of Physics \& Astronomy, University of Minnesota, Minneapolis, MN 55455, USA} 
\author{K.A.~McCarthy} \affiliation{Department of Physics, Massachusetts Institute of Technology, Cambridge, MA 02139, USA} 
\author{N.~Mirabolfathi} \affiliation{Department of Physics and Astronomy, and the Mitchell Institute for Fundamental Physics and Astronomy, Texas A\&M University, College Station, TX 77843, USA} 
\author{R.A.~Moffatt} \affiliation{Department of Physics, Stanford University, Stanford, CA 94305, USA} 
\author{S.M.~Oser} \affiliation{Department of Physics \& Astronomy, University of British Columbia, Vancouver,  BC  V6T 1Z1, Canada} 
\author{K.~Page} \affiliation{Department of Physics, Queen's University, Kingston ON, Canada K7L 3N6} 
\author{W.A.~Page} \affiliation{Department of Physics \& Astronomy, University of British Columbia, Vancouver,  BC  V6T 1Z1, Canada} 
\author{R.~Partridge} \affiliation{SLAC National Accelerator Laboratory/Kavli Institute for Particle Astrophysics and Cosmology, Menlo Park 94025, CA} 
\author{M.~Pepin} \affiliation{School of Physics \& Astronomy, University of Minnesota, Minneapolis, MN 55455, USA} 
\author{A.~Phipps} \affiliation{Department of Physics, University of California, Berkeley, CA 94720, USA} 
\author{K.~Prasad} \affiliation{Department of Physics and Astronomy, and the Mitchell Institute for Fundamental Physics and Astronomy, Texas A\&M University, College Station, TX 77843, USA} 
\author{M.~Pyle} \affiliation{Department of Physics, University of California, Berkeley, CA 94720, USA} 
\author{H.~Qiu} \affiliation{Department of Physics, Southern Methodist University, Dallas, TX 75275, USA} 
\author{W.~Rau} \affiliation{Department of Physics, Queen's University, Kingston ON, Canada K7L 3N6} 
\author{P.~Redl} \affiliation{Department of Physics, Stanford University, Stanford, CA 94305, USA} 
\author{A.~Reisetter} \affiliation{Department of Physics, University of Evansville, Evansville, IN 47722, USA} 
\author{Y.~Ricci} \affiliation{Department of Physics, Queen's University, Kingston ON, Canada K7L 3N6} 
\author{H.E.~Rogers} \affiliation{School of Physics \& Astronomy, University of Minnesota, Minneapolis, MN 55455, USA} 
\author{T.~Saab} \affiliation{Department of Physics, University of Florida, Gainesville, FL 32611, USA} 
\author{B.~Sadoulet} \affiliation{Department of Physics, University of California, Berkeley, CA 94720, USA}\affiliation{Lawrence Berkeley National Laboratory, Berkeley, CA 94720, USA} 
\author{J.~Sander} \affiliation{Department of Physics, University of South Dakota, Vermillion, SD 57069, USA} 
\author{K.~Schneck} \affiliation{SLAC National Accelerator Laboratory/Kavli Institute for Particle Astrophysics and Cosmology, Menlo Park 94025, CA} 
\author{R.W.~Schnee} \affiliation{Department of Physics, South Dakota School of Mines and Technology, Rapid City, SD 57701, USA} 
\author{S.~Scorza} \affiliation{Department of Physics, Southern Methodist University, Dallas, TX 75275, USA} 
\author{B.~Serfass} \affiliation{Department of Physics, University of California, Berkeley, CA 94720, USA} 
\author{B.~Shank} \affiliation{Department of Physics, Stanford University, Stanford, CA 94305, USA} 
\author{D.~Speller} \affiliation{Department of Physics, University of California, Berkeley, CA 94720, USA} 
\author{D.~Toback} \affiliation{Department of Physics and Astronomy, and the Mitchell Institute for Fundamental Physics and Astronomy, Texas A\&M University, College Station, TX 77843, USA}
\author{S.~Upadhyayula} \affiliation{Department of Physics and Astronomy, and the Mitchell Institute for Fundamental Physics and Astronomy, Texas A\&M University, College Station, TX 77843, USA} 
\author{A.N.~Villano} \email{Corresponding author: villaa@physics.umn.edu}\affiliation{School of Physics \& Astronomy, University of Minnesota, Minneapolis, MN 55455, USA} 
\author{B.~Welliver} \affiliation{Department of Physics, University of Florida, Gainesville, FL 32611, USA} 
\author{J.S.~Wilson} \affiliation{Department of Physics and Astronomy, and the Mitchell Institute for Fundamental Physics and Astronomy, Texas A\&M University, College Station, TX 77843, USA} 
\author{D.H.~Wright} \affiliation{SLAC National Accelerator Laboratory/Kavli Institute for Particle Astrophysics and Cosmology, Menlo Park 94025, CA} 
\author{X.~Yang} \affiliation{Department of Physics, University of South Dakota, Vermillion, SD 57069, USA} 
\author{S.~Yellin} \affiliation{Department of Physics, Stanford University, Stanford, CA 94305, USA} 
\author{J.J.~Yen} \affiliation{Department of Physics, Stanford University, Stanford, CA 94305, USA} 
\author{B.A.~Young} \affiliation{Department of Physics, Santa Clara University, Santa Clara, CA 95053, USA} 
\author{J.~Zhang} \affiliation{School of Physics \& Astronomy, University of Minnesota, Minneapolis, MN 55455, USA}

\smallskip
\date{\today}

\collaboration{SuperCDMS Collaboration}

\noaffiliation


\smallskip

\begin{abstract}
\cdmstwo{} data from the five-tower runs at the Soudan Underground Laboratory were reprocessed
with an improved charge-pulse fitting algorithm.  Two new analysis techniques to reject
surface-event backgrounds were applied to the 612\,\kgd{} germanium-detector weakly interacting
massive particle (WIMP)-search exposure.  An extended analysis was also completed by decreasing
the 10\,\keVr{} analysis threshold to $\sim$5\,\keVr{}, to increase sensitivity near a WIMP mass
of 8\,\gev{}.  After unblinding, there were zero candidate events above a deposited energy of
10\,\keVr{} and six events in the lower-threshold analysis. This yielded minimum WIMP-nucleon
spin-independent scattering cross-section limits of 1.8$\times$10$^{-44}$ and
1.18$\times$10$^{-41}$\,cm$^2$  at 90\% confidence for 60 and 8.6\,\gev{}  {}WIMPs, respectively.
This improves the previous \cdmstwo{} result by a factor of 2.4 (2.7) for 60 (8.6)\,\gev{}
{}WIMPs.  

\end{abstract}


\maketitle

\section{\label{sec:intro}Introduction}
The mass balance of the Universe is the subject of intense research and debate.  Discrepancies
between gravitationally determined galaxy-cluster masses and their observed luminosities provided
the earliest motivation for dark matter~\cite{1933AcHPh_6_110Z,*Zwicky2008,Babcock1939}.  Modern
galactic rotation curves sharpen the argument~\cite{Sofue2001,Begeman1991,Albada1986}, as do more
recent studies of galaxy cluster dynamics~\cite{Maughan2004,Lewis2003}.   Likewise, spectroscopy
of intergalactic x-ray-emitting gas~\cite{Maughan2004,Lewis2003} and gravitational
lensing~\cite{Kneib1996,Kneib1995,Sheldon2009,Sheldon2009a,Johnston2007} elevate these presumed
mass discrepancies to the level of a crisis.  

To bring the data sets into consistency requires either modifications of
gravity~\cite{Bekenstein2004,Milgrom1983} and/or large quantities of nonluminous matter.
Observations of colliding clusters~\cite{Kahlhoefer21012014,Clowe2006} provide evidence for excess
nonluminous matter, though alternate models of gravity and some nonluminous matter can apparently
also reproduce such results~\cite{Angus2007,Angus2006}.  Nonrelativistic (cold) relic particle
dark matter alone could resolve these discrepancies and is also considered an essential ingredient
in gravitational simulations of the large-scale structure of the Universe.  For example, the Via
Lactea and Millennium simulations show excellent agreement with the observed large-scale structure
of our Universe when cold dark matter is
included~\cite{Zemp:2009,Diemand:2011,Boylan-Kolchin21092009}.

While the observed galactic dynamics and large-scale structure naturally lead us to consider cold
particle dark matter, cosmological measurements have been important in constructing a consistent
model for the evolution of the Universe using such cold dark matter (CDM).  The accelerating
expansion of the Universe~\cite{Riess2004,perlmutter1999,Riess1998}, big bang
nucleosynthesis~\cite{Cyburt2008}, baryon acoustic oscillations~\cite{Eisenstein2005}, and the
cosmic microwave background~\cite{Ade:2013ktc,Hinshaw2012} support a cosmology whose dominant
components are dark energy (which could correspond to a cosmological constant $\Lambda$) and
nonbaryonic CDM.  When interpreted within the framework of the $\Lambda$CDM model, these
cosmological measurements enable precise determination of the CDM and dark-energy content of the
Universe~\cite{Kowalski2008}.

Particle physics provides clues as to the possible identity of nonbaryonic CDM.  It was realized
early on that weakly interacting massive particles (WIMPs) with GeV- to TeV-scale masses could
thermally freeze out in the early Universe to give the correct relic density~\cite{Lee1977}.
Supersymmetry provides a WIMP candidate in the lightest supersymmetric particle and has many other
benefits, like solving the hierarchy problem~\cite{Aitchinson2007}.  Still other particle physics
models considered more recently~\cite{Zurek2009} provide motivation for light WIMPs and linkages
to the matter/antimatter asymmetry of the Universe.

WIMPs can be searched for directly through their scattering off nuclei in a terrestrial detector.
Since these interactions are expected to be rare it is important for a dark-matter detector to
have a low threshold (10\,\keVr{} or below) and excellent background rejection capabilities.  The
Cryogenic Dark Matter Search (CDMS) collaboration has developed cryogenic semiconductor detectors
focusing on those properties for the purpose of measuring WIMP-scattering events. 

\begin{figure}[!htb]
   \includegraphics[width=\columnwidth]{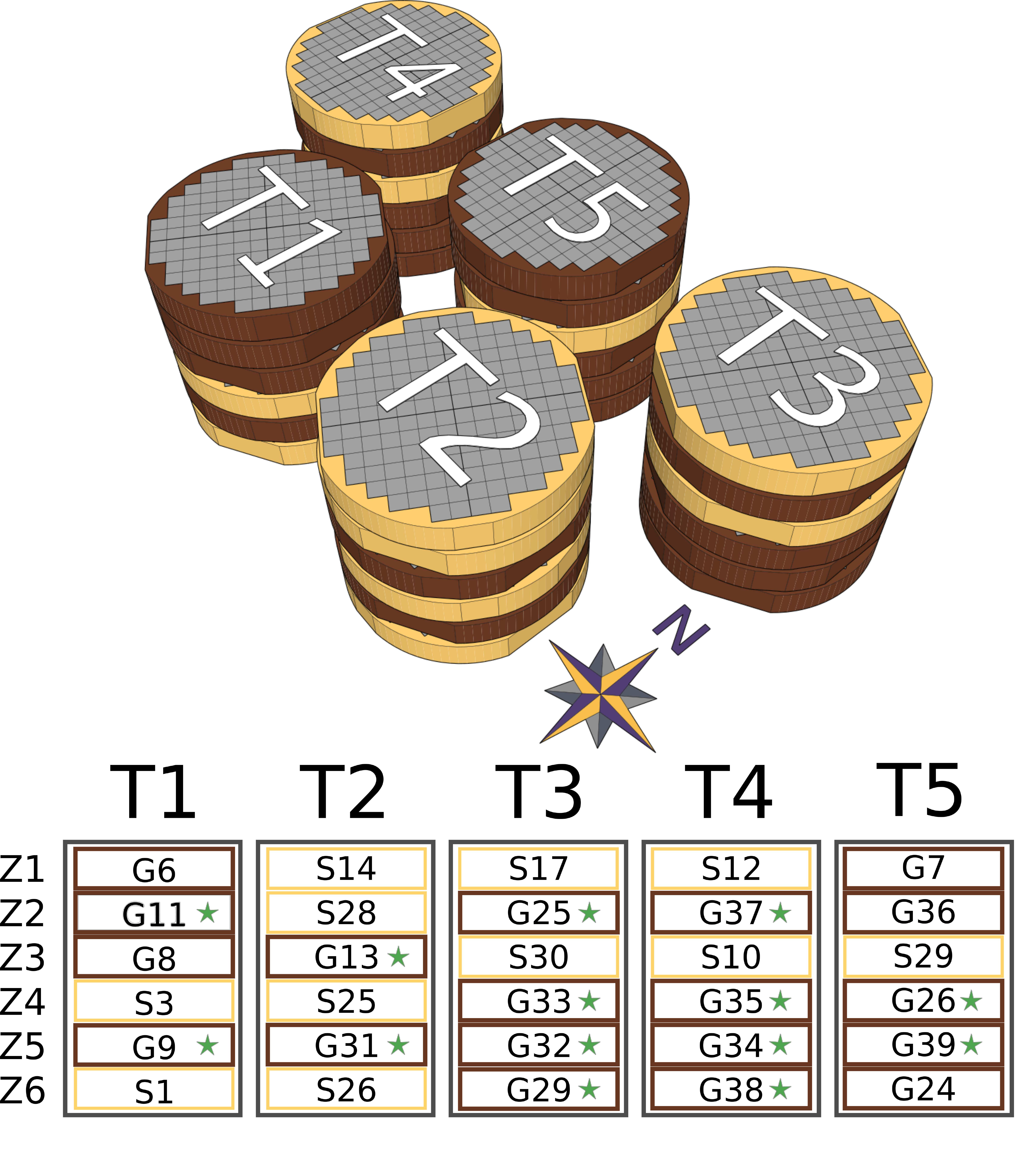}
   \caption{\label{fig:stack}(Color online) Isometric representation of the \cdmstwo{} detector
   arrangement and tower occupation, with direction of north indicated.  The ZIP detectors in each
   tower are numbered 1--6 and are either silicon (yellow) or germanium (brown).  The
   identification numbers of each detector were given to track the raw material the detectors were
   produced from and the processing to which they were subjected.  For example the detector ``G9''
   is a germanium detector with identification number 9.  Green stars indicate detectors that were
   used in this data analysis (see \sect~\ref{sec:det_princ}). 
   }
\end{figure}

This paper explores new analysis techniques using data taken by the CDMS experiment for the direct
detection of WIMPs during the \cdmstwo{} running
period~\cite{Ahmed2010a,Ahmed2011,CDMS_EDEL_2011}.  There are four data periods associated with
the data set used in this work, varying from 1 to 6\,months in duration.  The experiment used Ge
and Si detectors and was located in the Soudan Underground Laboratory at a shielding depth of 2090
meters water equivalent (m.w.e.).  The largest payload consisted of a total of 19 Ge ($\sim$240\,g
each) and 11 Si ($\sim$110\,g each) detectors.  All detectors are cylindrical, 7.6\,cm in
diameter, $\sim$1\,cm in height, and arranged in five vertical stacks (``towers'') each including
six detectors~\cite{Ahmed2009,Akerib2005} (see \fig~\ref{fig:stack}).  The detectors are labeled
T$x$Z$y$ where $x$~(1--5) is the tower number and $y$~(1--6) indicates the position within the
stack (from top to bottom). 

In each detector, particle interactions produce electron-hole pairs (ionization) together with
phonons.  The charge carriers are drifted by a small electric field (3\,V/cm for Ge and 4\,V/cm
for Si) and collected on concentric aluminum electrodes deposited on one of the flat faces of the
crystal.  On the opposite face, superconducting transition-edge sensors (TESs) arranged in
quadrants collect phonons before thermalization.  Charge and phonons are measured independently
for the purpose of event-by-event discrimination:  WIMP signal events will produce a nuclear
recoil (\nr) in the detector, whereas most background processes produce an electron recoil (\er).
The ratio of ionization to phonon signal amplitudes allows discrimination of \nr{s} from the far
greater number of \er{s} with a rejection factor $>$\,10$^4$~\cite{Ahmed2010a} in the
10--100\,\keVr{} region.  Lead shielding further reduced gamma-induced backgrounds, and
neutron-induced \nr{} backgrounds were reduced with polyethylene shielding.  The shielding is
surrounded by an active scintillator veto to tag events induced by cosmic-ray muon showers.  

Recoils within $\sim$10\,$\mu$m of the detector surface can have reduced ionization signals
because of poor charge collection, which can be sufficient to misclassify a surface \er{} as an
\nr{}.  These reduced-charge \er{s} originate from three major sources:  electrons produced by
$\beta$ emitters on or near the detector surfaces, electrons ejected from photons scattering in
nearby material, and photons that interact near the detector surfaces.  All of these mechanisms
can lead to events in which an energy deposition is observed in only a single detector, producing
``\seuh{s}'' that can be mistaken for single-interaction \nr{s}.  The long-lived $^{222}$Rn
daughter $^{210}$Pb, implanted in the detector surfaces and their copper housings, is the primary
source of such surface events.  This class of events constitutes the dominant background for the
\cdmstwo{} WIMP search and presents a considerable challenge~\cite{Akerib2005}.  Fortunately,
recoils near detector surfaces are characterized by prompt phonon absorption in the TESs.
Consequently, the phonon signals for surface \er{s} are (on average) faster than for \nr{s} in the
detector's bulk, enabling \se{} background discrimination based on phonon-pulse timing.   The
event selection criteria derived from phonon timing (called ``timing cuts'') presented here are
essential to obtaining optimal WIMP sensitivity for the \cdmstwo{} data set because they mitigate
the \se{} background, improving the overall \er{} rejection to $>$\,10$^6$. 

Nonsignal \nr{} events have two known sources: Ge recoils induced by scattering neutrons and
$^{206}$Pb nuclei from the decay of $^{210}$Po near the detector surfaces.  The expected neutron
background for \cdmstwo{} is roughly equal parts cosmic-ray muon-induced neutrons and radiogenic
neutrons from trace contaminants in the shielding and detector.  Radiogenic neutrons produced
outside the shielding have too little energy to cause a detectable \nr{} after penetrating the
polyethylene.  By combining simulations with \emph{in situ} data, we estimate the neutron
background to be subdominant compared to the \se{} background (see
\psect~\ref{sec:cosneutron_bknds} and~\ref{sec:radneutron_bknds}).  The $^{206}$Pb background
contribution is also subdominant in this analysis.

The data set used in this work comes from the Ge detectors in the final \cdmstwo{} five-tower
exposure and was acquired between July 2007 and September 2008.  The total raw exposure for this
running period was approximately 612\,\kgd.  The original analysis of this data set provided
world-leading sensitivity to spin-independent elastic WIMP-nucleon scattering in 2010 when it was
first published~\cite{Ahmed2010a}.  Here, we reevaluate this data set using improved data
reduction algorithms and \se{} rejection methods that reduce the expected \se{} contamination
(``leakage'') in the WIMP signal region compared with the original analysis.  In the 2010 analysis
a 10\,\keVr{}~\footnote{\label{foot:kev}  Throughout this work the unit keV shall refer to true
recoil energy whereas keV$_\textrm{ee}$ are units of electron-equivalent recoil energy.  This
means that an ionization yield $y = E_q / E_r = 1$ is assumed when the energy is calculated from
the measured signal.} analysis threshold was used to limit the expected background to less than
one event for the entire exposure.  Intriguing results at low WIMP
mass~\cite{Agnese2013,Aalseth2013,Angloher2012,Aalseth2011,Bernabei2010} motivated us additionally
to examine the data with reduced thresholds.  

\section{\label{sec:det_princ}\cdmstwo{} Detector Properties}
\begin{figure}[!htb]
   \includegraphics[width=\columnwidth]{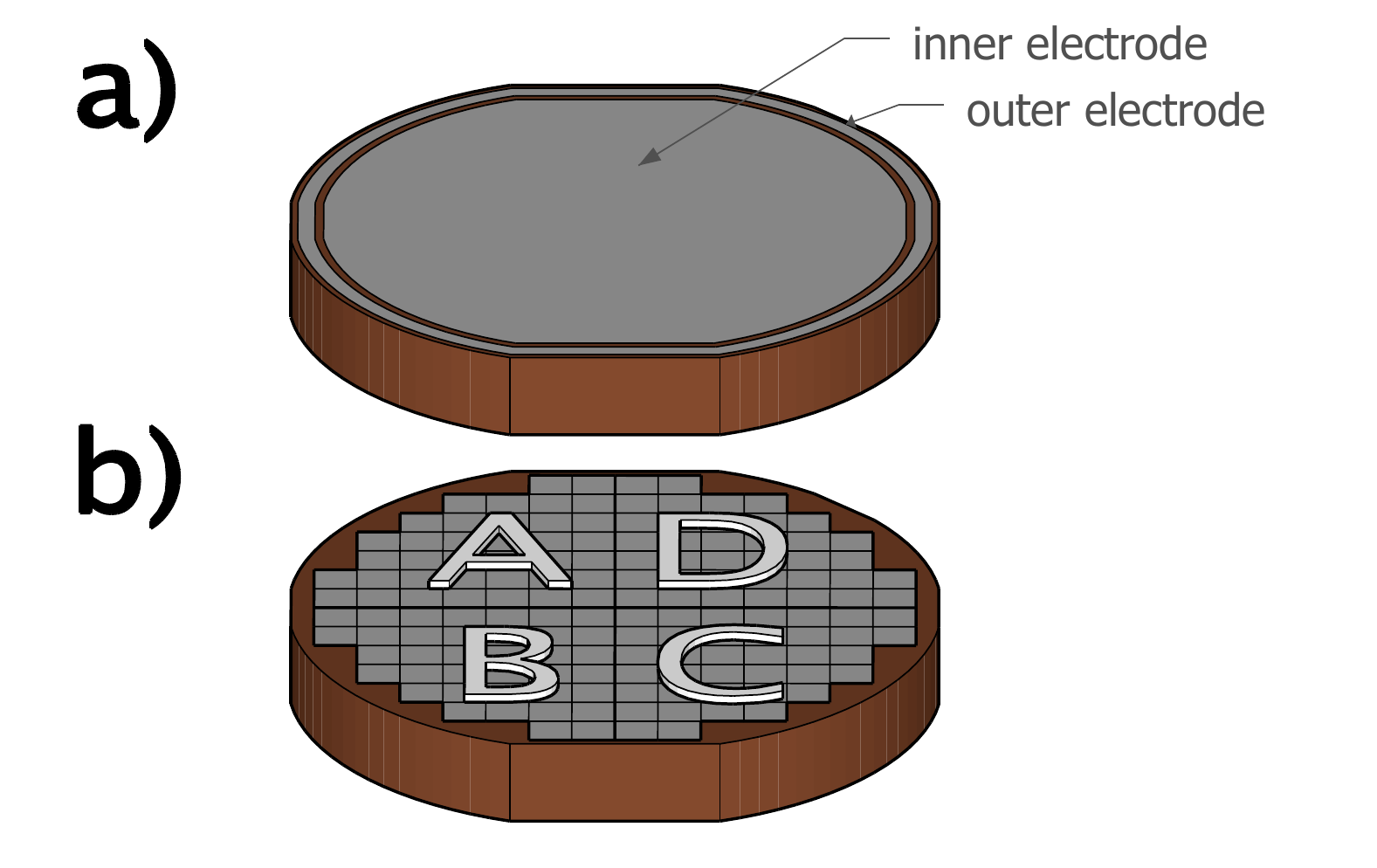}
   \caption{\label{fig:zip}(Color online) a) Inner and outer electrodes for measuring the
   ionization signal.  The inner electrode extends from the center to a radius of 34.5\,mm and the
   outer electrode extends from a radius of 35.5\,mm to just before the edge (0.5--2\,mm; the
   diameter of the detectors is 76\,mm).  b) Phonon channels (QETs) A--D arranged in quadrants on
   the opposite side of the detector.  Each phonon channel is composed of approximately 1000 TESs
   formed from thin films deposited onto the crystal surface and then photolithographically
   structured.  The TESs are connected in parallel to form the QETs and each QET is read out by
   its own circuit containing a SQUID array.
   }
\end{figure}

The \cdmstwo{} Z-sensitive ionization and phonon detectors (ZIPs) are operated at a temperature of
$\sim$50\,mK and feature six readout channels:  two charge electrodes on one side and four phonon
sensors on the opposite side~\cite{Hellmig2000}.  The analysis undertaken here includes only Ge
detectors, for which the ionization channels are biased with a 3\,V potential across the crystal.
Furthermore, five of the 19 Ge detectors were omitted from the analysis because of readout channel
failures across all data sets.  This gives a direct analogy to the original analysis of this data
set, which used the same detector subset.  Ionization channels are read out by a low-noise
junction field-effect transistor circuit with an operating temperature of
$\sim$150\,K~\cite{Akerib2008476}.  Phonons are detected by quasiparticle-trap-assisted
electrothermal-feedback transition-edge sensors (QETs). The QET signal is amplified by
superconducting quantum-interference devices (SQUIDs) that are thermally coupled to the cryostat's
600\,mK cold stage~\cite{Akerib2008476}.  Figure~\ref{fig:zip} shows a schematic of the ZIP
channel layout.  The outer charge electrode acts as a veto against events that deposit energy near
the sidewalls, where higher background is expected and charge collection is more likely to be
incomplete.  

Data taken during the \cdmstwo{} experiment are composed of calibration (using a $^{133}$Ba gamma
source, or a $^{252}$Cf neutron source) and WIMP search.  The $^{133}$Ba calibration data were
used to determine the energy scale and to study various systematic effects.  The $^{252}$Cf
calibration data were used to study detector response to \nr{s} and act as a proxy for determining
WIMP acceptance.

The \cdmstwo{} front-end electronics issued experimental triggers in response to activity in
either the surrounding scintillator veto or the ZIPs.  For veto-triggered events, the front-end
electronics required two or more veto panels to have coincident signals in excess of their
hardware thresholds.  ZIP-triggered events occurred when a composite phonon signal for any ZIP
exceeded its preset discriminator threshold.  The composite phonon signal was the analog sum of
the four phonon signals from the ZIP with a 900--18000\,Hz band-pass filter applied to reduce
noise-like fluctuations.  During $^{133}$Ba calibration runs, selective readout was employed
because of the high rate; i.e.\ only detectors in the tower in which a ZIP trigger occurred were
read out.  During $^{252}$Cf calibration and WIMP-search runs, all detectors were read out in
response to either veto or ZIP triggers.  Veto-triggered WIMP-search events were recorded for use
in studies of cosmogenically induced neutron backgrounds.  Each triggered event includes two charge
and four phonon traces (per ZIP readout).  Each trace consists of 2048 digitized amplitudes acquired
at a rate of 1.25\,MHz, with 512 samples prior to the trigger time.

Charge carriers moving in the electric field of the detector generate phonons with a total energy
proportional to the potential difference traversed (Neganov-Luke
effect~\cite{Neganov:1985,Luke:1988}). These ``Luke phonons'' are typically ballistic and add to
the phonon signal generated by the primary recoil.  The recoil energy $E_r$ was constructed from
the total phonon energy $E_p$ by subtracting a term proportional to the ionization-derived recoil
energy $E_q$ (or ``ionization energy'') to account for the Neganov-Luke effect:
\begin{equation}\label{E:rec_en}
E_r = E_p - \frac{eV}{\epsilon} E_q, 
\end{equation}
where $e$ is the elementary charge, $V$ is the absolute value of the operating potential of the
detector, and $\epsilon$ is the average electron-hole pair creation energy.  In Ge the value of
$\epsilon$ is approximately 3.0\,eV/pair for \er{s}~\cite{Alig:1975}.  The ``ionization yield''
(or yield) is defined as the ratio of ionization energy to recoil energy ($E_q/E_r$) and we
require this quantity to be unity for \er{s} (see below).  An \nr{} will produce less ionization
than an \er{} of equal recoil energy.  This effect is well known~\cite{Lewin1996} and provides the
basis for the ZIP detector's primary method of \er{}/\nr{} discrimination.  For recoil energies
between 10 and 100\,\keVr{} in Ge, the \nr{} ionization yield varies between 0.2 and 0.3.

Using the $^{133}$Ba calibration data, the charge and phonon amplitudes were calibrated so that
the energies of known gamma lines are reproduced and the yield is unity for \er{s}.  Specifically,
the charge channels were calibrated using the 356\,keV line after a $\sim$10\% correction to the
ionization amplitudes to account for small systematic variations with interaction location (within
the detector).  We do not fully understand the origins of the systematic variation with
interaction location, though it is empirically robust.  One possibility is that the detectors are
not neutralized uniformly by the infrared (940\,nm) LED that is activated between data-taking
periods to remove trapped charge.  The relative phonon-channel calibration was performed by
scaling the phonon-channel amplitude fractions--the amplitudes of individual channels divided by
the sum of the amplitudes--to have equivalent distributions.  Finally, the summed phonon energy
was calibrated by requiring the ionization yield to be unity on average for \er{} events in the
65--100\,\keVr{} region.

\section{\label{sec:reprocessing}Raw Data Reduction and the Reprocessing}
Our analysis parameters are calculated from digitized charge and phonon pulses for each ZIP
detector using a set of pulse-processing algorithms.  These algorithms distill timing and
amplitude information from the four phonon and two charge traces (per ZIP).  Surface events are
rejected using timing quantities derived from the raw data, such as the rise time $\tau$ and the
delay $t_{\mathrm{del}}$ of the largest of the four phonon pulses with respect to the faster
ionization signal.  The amplitude of the inner charge-channel pulse gives a measurement of $E_q$
and the amplitude of the analog sum of the four phonon pulses gives a measurement of $E_p$ after
calibration.  These energy variables are used to construct the ionization yield ($y$) and recoil
energy ($E_r$) described in \eq~(\ref{E:rec_en}).  

The relative charge-channel amplitudes (charge ``partition''), relative phonon-channel amplitudes
(phonon ``partition''), and relative phonon pulse timing provide information about an event's
position within the detector.  These additional parameters are also calculated using the output of
the pulse-processing algorithms and are used for the phonon event-position-based correction and
charge-derived fiducial-volume restrictions.  

This section gives an overview of the most important algorithms for constructing these analysis
parameters. It also explains the upgrade to the charge-pulse processing that motivated the
reprocessing of this data set.

\subsection{\label{sec:par_extraction}Parameter extraction}
\textbf{Timing parameters}.  Timing estimators are derived using an algorithm that steps along a
low-pass-filtered trace to identify the pulse rise time (RT) and fall time (FT).  We call this the
RT-FT-walk algorithm.  The traces are first filtered with a low-pass Butterworth
filter~\cite{Butterworth1930}.  Two sets of parameters are produced; one where the cutoff
frequency is 50\,kHz and one where it depends on the signal-to-noise of the trace.  The filter
removes high-frequency noise and effectively smooths the pulse for an improved determination of
the RT and FT at various percentages of the pulse maximum.  RT and FT information is determined by
``walking'' along the filtered trace starting at the maximum and identifying the times at which
the respective threshold levels are reached.  For example, it infers the time at which the rising
pulse edge reaches 20\% of the pulse's maximum amplitude.  The RT-FT-walk algorithm was applied to
all charge and phonon traces.  The rise time $\tau$ is computed as the 10--40\% time span along
the rising edge of the largest of a detector's four phonon pulses using the RT-FT-walk algorithm.

\textbf{Optimal filtering}.  A pulse-template optimal filter (OF)~\cite{press_numerical_1992} is
used to produce the best resolved energy quantities.  The OF has superior energy resolution
compared to pulse-integral quantities and is well suited to the analysis of small recoil
energies~\cite{SunilThesis}.  A template for the expected pulse shape is fit to the pulse in
Fourier space, deweighting the frequency bins with high noise.  The frequency deweighting is done
for each individual data ``series'' -- a data-taking block normally lasting between 10 and
12\,h--using noise power spectral densities (PSDs).  The PSD is constructed from randomly
triggered traces taken before the series to sample the noise environment.  Phonon pulse templates
are two-exponential functional forms, with rise- and fall-time parameters tuned to match
individual detectors by fitting to an average pulse.  Charge pulse templates were produced
empirically by averaging normalized data traces.  This fitting procedure is done for all possible
pulse template delays and the best-fit delay is chosen.  

A single-pulse-template OF is performed on each phonon pulse separately and on the sum of a
detector's four phonon traces.  For charge pulses, to account for crosstalk between the inner and
outer channels, the two pulses are fit simultaneously with a crosstalk-correcting OF (OFX)
described in more detail below.  During the OF fit an array of best-fit amplitudes is produced
that corresponds to each possible delay of the template with respect to the experimental pulse.
The maximum such amplitude in a preselected delay window is chosen and its time defines the delay
quantity.  For an OF using a single pulse template this maximum amplitude choice also has the
lowest $\chi^2$ in the preselected window.  The OF amplitude for the summed phonon trace gives
$E_p$ after calibration and the inner charge amplitude from the OFX procedure gives $E_q$ after
calibration.  The charge-to-phonon delay $t_{\mathrm{del}}$ is determined from the difference
between the OFX charge delay and the 20\% crossing of the largest phonon signal as determined by
the RT-FT-walk algorithm.  The delay is computed as follows:
\begin{equation}\label{E:tdel}
t_{\mathrm{del}} = t_{p20} - t_{\mathrm{OFX}}, 
\end{equation}
with $t_{\mathrm{OFX}}$ being the OFX ionization delay relative to the global trigger time and
$t_{p20}$ being the 20\% crossing time of the largest phonon signal relative to the global
trigger.

\subsection{\label{sec:par_corrections}Parameter corrections}
\textbf{Phonon position correction}.  The phonon sensor response is highly position
dependent~\cite{Mandic2002}.  In order to optimize the derived phonon variables, a position
correction was used to modify energy, timing, and yield quantities by normalizing to the mean of
nearby events.  We first defined averaging neighborhoods using a five-dimensional metric that
included total phonon energy, two relative phonon delay parameters, and two phonon energy
partition parameters~\cite{ZeeshThesis}.  For each parameter to be corrected,  a lookup table
containing the neighborhood averages for each five-dimensional bin was then constructed using bulk
\er{} events confirmed to be of good quality from $^{133}$Ba calibration.  To apply the correction
we created a corrected version ($v_c$) of a given parameter ($v$) as in \eq~(\ref{E:poscorr}):
\begin{equation}\label{E:poscorr} 
v_c = \frac{v}{\langle v \rangle_{\mathrm{bin}}} \langle v \rangle_{\mathrm{global}}, 
\end{equation} 
where the subscript ``global'' refers to averaging over the whole calibration data set, and
``bin'' refers to averaging over one neighborhood.  After the lookup table was applied to the
$^{133}$Ba calibration data, the selection of bulk \er{s} was refined to create the final
correction table, which was then applied to the WIMP-search data.  The final energy, timing and
yield quantities were made more uniform across the detector for bulk events by using this
procedure, thus improving the rejection of \seuh{s}.

\textbf{Charge crosstalk}.  Charge signal quantities use the OFX algorithm to account for
crosstalk. Using separate pulse-template OFs for the inner and outer electrodes, a single delay
was chosen.  In the original analysis of our data set~\cite{Ahmed2010a}, this delay corresponded
to the delay that made the sum of the two OF amplitudes maximal.  While this is likely to be very
close to the delay that minimizes the $\chi^2$ of the fit, it is not guaranteed because of the
combined fitting of the inner and outer electrodes.

\subsection{\label{sec:re_motivation}Reanalysis motivation}
\begin{figure}[!htb]
\begin{center}
  \includegraphics[width=\columnwidth]{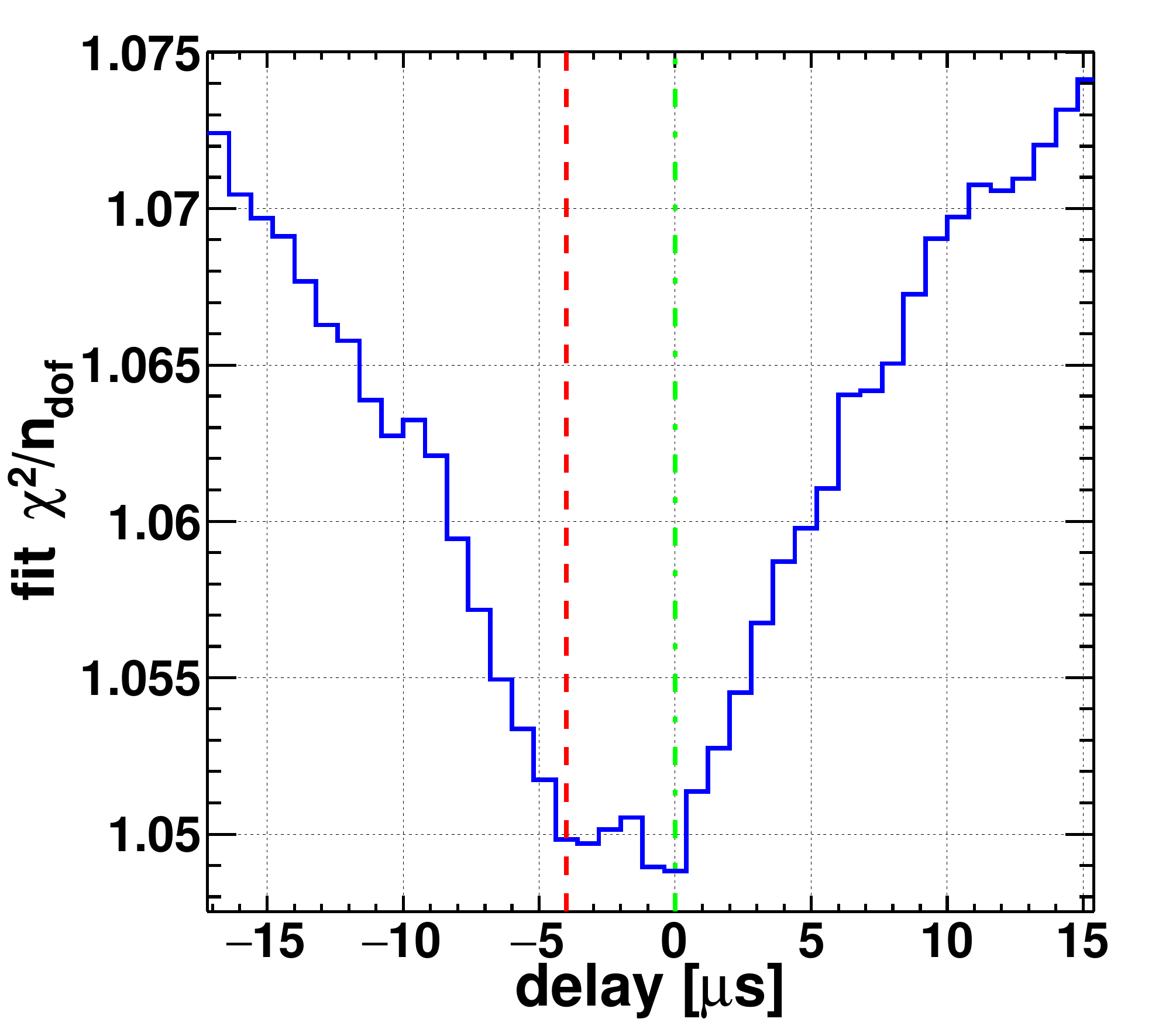}
  \caption{\label{fig:candi_chi} (Color online) The OFX $\chi^2$ as a function of delay for the
  T3Z4 WIMP candidate from the original analysis~\cite{Ahmed2010a}.  The delay chosen by the
  original ionization fit to this event (red dashed line) is not a global minimum in $\chi^2$,
  rather it is displaced $\sim$4\,$\mus$ from the global minimum (green dot-dashed line). While
  this small error in the inferred delay for this and other events did not affect the measured
  energy significantly (a $\sim$2\% decrease), it caused an error in the \se{-}rejection timing
  parameter for low-energy events.  
  }
\end{center}
\end{figure}

After finalizing the first analysis of this data set in 2009~\cite{Ahmed2010a} we noticed that one
of the WIMP candidate events had an OFX delay value that did not correspond to a global minimum in
the $\chi^2$ of the fit (see \fig~\ref{fig:candi_chi}).  This issue was one of the main reasons
for proceeding with a reanalysis.

The reason for implementing the maximum-amplitude method instead of a full $\chi^2$ minimization
in the original analysis was to save computing time; a global trigger issued by one detector
forces the readout of all detectors, giving a substantial number of traces with near-zero
amplitudes that need not be processed with the time-consuming $\chi^2$ minimization.  The result
of using this maximum-amplitude method is to smear the ionization-to-phonon delay distributions
slightly, which can cause additional background events to leak past the timing cuts.  This delay
smearing affects lower-ionization events more than higher-ionization events.  In particular,
\nr{s} that deposit energy in a single detector (``single scatters'') are more susceptible than
the (higher-ionization) low-yield \er{s} used to quantify the expected leakage of \seuh{s} into
the signal region.  

The OFX inaccuracy made only a small contribution to the previously published limit, and
background estimates were adjusted upward in the original result to account for the resulting
larger uncertainties at low energies~\cite{Ahmed2010a}.  Nevertheless, the number and character of
the candidate events provides useful information on the possibility of a signal.  It is very
likely that the candidate event featured in \fig~\ref{fig:candi_chi} would have been removed by
the timing cut in the original work had the ionization delay at the global minimum of the $\chi^2$
been chosen, but using the improved OFX procedure also has the potential to cause previously
excluded events to become WIMP candidates.  

For the reasons stated above, the data were reprocessed, selecting the global minimum of the
$\chi^2$ in the OFX algorithm for most ionization pulses, rather than the maximum-amplitude
method.  Because of the large number of traces in the raw data consistent with noise, traces with
pulses corresponding to charge energies below a detector-dependent threshold
(0.94\,keV$_{\mathrm{ee}}$~\cite{Note1} on average across the 14 detectors used in this work) were
still processed with the maximum-amplitude method in order to save processing time.  This has no
effect on the results presented here because this energy is below our lowest analysis threshold.

\section{\label{sec:data_sel_eff}Data Selection and Efficiencies}
Once the data set was calibrated and position corrected, we created several selection criteria (or
``cuts'') to produce the cleanest signal-region sample possible.  Since the rejection of
nonsignal events is typically not perfect, a signal event retention efficiency was computed for
each cut.  The cuts and efficiencies are covered in this section, starting with the efficiency for
the trigger--a selection criterion that is made in hardware before the events are recorded.  The
result is a signal-region sample corresponding to a well-known exposure. 

\begin{figure}[!htb]
	\includegraphics[width=\columnwidth]{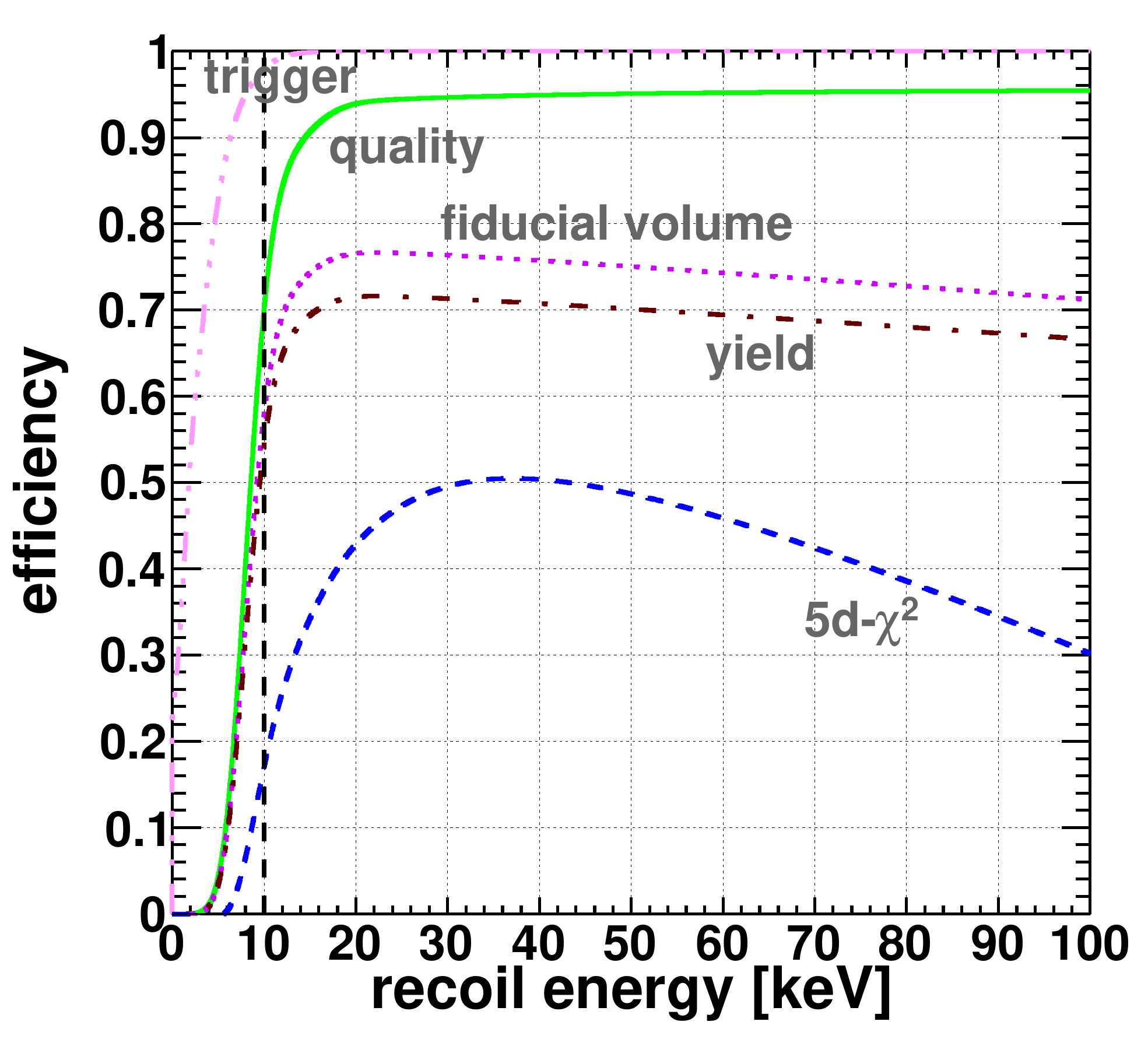}
	\caption{\label{fig:c58r_combeff_chisq_58} (Color online) Detection efficiency combined
	across all detectors as a function of energy for different classes of cuts.  The curves
	show the total efficiency as more cuts are added beyond the hardware trigger (pink
	triple-dot-dashed): event-level data-quality cuts (green solid), ionization-based
	fiducial-volume cut (magenta dotted), ionization-yield cut (maroon dot-dashed), and
	\fivedchi{} (see \sect~\ref{sec:5dchi_timing}) phonon timing cut (blue dashed).  The
	falloff of the data-quality cut efficiency below $\sim$20\,\keVr{} is due to the interplay
	between the ionization threshold and the requirement that the ionization yield be
	3$\sigma$ below the \er{} band (see main text).  The vertical black dashed line is our
	standard 10\,\keVr{} analysis threshold.  
	}
\end{figure}

A trigger efficiency for each detector was determined as a function of phonon energy and converted
to a function of recoil energy using our measured \nr{} ionization yield from $^{252}$Cf data.   A
given detector's trigger efficiency was calculated using all WIMP-search events in which another
detector caused an experimental trigger.  When the global trigger initiates a readout of all the
detectors, other delayed instances where phonon pulses were in excess of the corresponding
discriminator thresholds were recorded into a logical buffer.  Regardless of the content of these
trigger records, for each event and detector the optimal-filtering techniques described in
\sect~\ref{sec:det_princ} were used to reconstruct the total phonon energy ($E_p$) from
\eq~(\ref{E:rec_en}).  The trigger efficiency curve is then the ratio of two recoil-energy
spectra: the spectrum of events with a phonon trigger in both the detector in question and another
detector divided by the spectrum of events with a phonon trigger in another detector.  As shown in
\fig~\ref{fig:c58r_combeff_chisq_58}--combined across the 14 Ge detectors considered here--the
trigger efficiency is $\sim$100\% for recoil energies above our standard analysis threshold of
10\,\keVr{}.  Note that the extended analysis described in \sect~\ref{sec:timing_ex} requires the
trigger efficiency down to $\sim$5\,\keVr{}.

Events above the trigger threshold were subjected to five classes of cuts used to isolate
high-quality signal candidates in the WIMP-search data: cuts that remove time periods of reduced
overall data quality; event-level data-quality cuts; an ionization-based fiducial-volume cut to
reject events near the detector sidewall; an ionization-yield cut; and one of three phonon timing
cuts.  Cuts that remove time periods cause a loss of experimental live time because events are
removed uniformly for all detectors and in all kinematic variables.  After the live-time cut, the
total remaining live time for each detector was computed and stored.  All other cuts cause
reductions in \nr{} detection efficiency, some of which vary significantly as a function of recoil
energy.  We computed these ``efficiency functions'' and applied them to each detector in
conjunction with the live time to compute the final exposures.

The \nr{} detection efficiency functions were estimated in bins of recoil energy.  Well-motivated
functional forms were fit to the bin-wise estimates, with the best-fit results shown in
\fig~\ref{fig:c58r_combeff_chisq_58}.  Most cuts have little effect on the acceptance of nuclear
recoils.  The ionization-based fiducial-volume cut and the phonon timing cuts cause the greatest
loss of signal acceptance. Of the three timing cuts described in \sect~\ref{sec:timing_analysis},
use of the ``\fivedchi{}'' timing cut results in the highest final acceptance above 10\,\keVr{},
about 50\% at $\sim$35\,keV.  The \fivedchi{} timing-cut efficiency is shown in
\fig~\ref{fig:c58r_combeff_chisq_58} as an example while the other timing-cut efficiencies are
detailed in \sect~\ref{sec:final_timing_efficiency}.

\subsection{\label{sec:livetime_cuts}Live-time cuts}
Live time was removed during periods with disabled readout channels, poor detector neutralization
(characterized by increased levels of charge trapping in the crystal bulk~\cite{SunilThesis}),
decreased resolution, improper experimental configurations, and trigger anomalies that consist of
isolated bursts of events or incorrect phonon trigger threshold.  The Soudan Underground
Laboratory also houses a neutrino detector, MINOS, to measure properties of the ``NuMI'' neutrino
beam originating from Fermi National Accelerator Laboratory~\cite{Abramov2002209}.  While it is
very unlikely for CDMS to observe any beam-induced events, the loss of live time incurred by
removing all time periods coincident with a NuMI beam spill is negligible.  The NuMI neutrino-beam
cut is implemented as a restriction on the live time, removing events within 60\,$\mu$s of the
arrival of the neutrino beam's 10\,$\mu$s spills.  After application of these cuts, 612\,\kgd{} of
total exposure remain.  We define this as our ``raw exposure.''  Out of a total of 36\% loss of
live time (64\% data-taking efficiency) the largest contributors were bad environmental
configurations ($\sim$15\%), failed KS tests ($\sim$8\%), cryocooler noise ($\sim$6\%), poor
phonon reconstruction ($\sim$4\%), and fundamental hardware failures ($\sim$3\%).  Bad
environmental configurations can consist of many effects including trigger bursts, high charge
noise, or insufficient LED flashing. 

\subsection{\label{sec:quality_cuts}Quality cuts}
The event-level class of quality cuts consists of several components.  A ``glitch'' cut removes
events that have phonon pulses resembling electronic noise in the phonon readout chain.  These
events are typically correlated across multiple detectors and are characterized by phonon pulses
with fall times shorter than the time scale expected for phonon dissipation.  Such events are less
likely to appear in the ionization readout chain; consequently, they are effectively removed with
a cut that rejects events in which the phonon pulse multiplicity is significantly larger than the
ionization pulse multiplicity~\footnote{We reject events that have phonon pulse multiplicities at
least 4 larger than the charge pulse multiplicity.  Further, if the phonon pulse multiplicity is 3
we reject events with charge multiplicity 0.}.  The muon-veto cut uses the 2\,in. thick scintillator
panels that surround the \cdmstwo{} experiment and the following two rejection criteria: an event
was removed if 1) a ZIP event is associated with a muon-like energy deposition (0.58\,V or
$\sim$3.8\,MeV deposited) in any veto panel occurring between 185\,$\mus$ before and 20\,$\mu$s
after the ZIP trigger time, or 2) there is any veto activity above the scintillator-panel
hardware threshold ($\sim$0.23\,V) and within the 50\,$\mu$s before a ZIP trigger.  Both criteria
are used for the muon-veto cut because it makes the cut stronger and has only about a 2\%
efficiency loss across all recoil energies.  Finally, because WIMPs will not interact in more than
one detector within a given event, we defined a ``single-scatter'' cut to select events involving
a single ZIP detector as follows: 1) the phonon signal must be greater than six standard
deviations above the mean electronic-noise level in the detector under consideration; and 2) the
signal in all other detectors must be within four standard deviations of the means of their
respective noise distributions.  Ionization channels were used for multiple-scatter rejection for
those detectors with degraded phonon channel performance. Thus, Ge and Si detectors that are not
part of the WIMP-search exposure were still live with respect to identification and rejection of
multiple-scatter events.  The single-scatter cut efficiency for each detector was estimated from
the fraction of randomly triggered events (i.e., electronic noise) that satisfy criterion 2).
The combined efficiency of the glitch, muon-veto and singles cuts varies by detector and over
time, and ranges from 96\% to 98\%.  Another aspect of the singles cut is its use to select
certain samples from calibration data.  For example, multiple-scatter events that have their
secondary scatter in the detector above or below the triggering detector are said to be
``face tagged.'' For these events the primary recoil is biased toward the direction facing the
multiples tag.  When using the multiples in background estimations, separating the events by face
helps decrease the systematic uncertainties (see \sect~\ref{sec:surface_event}).  Sometimes events
with multiples toward the phonon readout side are called ``phonon-side'' and those with multiples
toward the charge readout side are called ``charge-side.''

Some quality cuts depend explicitly on the kinematic variables.  One of these is an ionization
threshold cut, which requires events to have reconstructed ionization signals greater than 4.5
standard deviations above the mean noise level.  This cut removes nearly all events with zero
charge collected (those from very near the side walls), resulting in $<$0.1 of such events over
the whole exposure.  A charge-pulse reconstruction-quality cut was also defined, requiring the
$\chi^2$ value of the OFX fit to be less than an energy-dependent threshold.  This cut selects
events with high-quality charge energy estimators, suppressing those with excess electronic noise
or the occurrence of multiple pulses within a single event trace (referred to as ``pile-up'').
Its efficiency was measured as a function of charge energy using ${}^{133}$Ba calibration data and
translated into recoil energy using the average \nr{} ionization yield measured from $^{252}$Cf
calibration.  Above $\sim$20\,keV$_{\textrm{ee}}$~\cite{Note1} the efficiency is constant and $>
99$\%.  It decreases slightly at lower recoil energies.

Charge collection is reduced near the cylindrical walls of the detector, causing lower ionization
yield and thus poor \nr{} to \er{} discrimination.  The ionization-based fiducial-volume cut
accepts only events with an outer electrode signal consistent with noise.  The efficiency of this
cut (see \fig~\ref{fig:c58r_combeff_chisq_58}) was estimated using the $^{252}$Cf calibration
data, including a small correction ($\sim$13\% maximum across the energy range) accounting for
residual leakage of \er{s} into the \nr{} signal region in these data.  Another $\sim$5\%
correction was applied across the whole energy range to account for detector self-shielding to
neutrons and multiple scattering; this correction was based on a Monte Carlo simulation of neutron
scattering.  

\subsection{\label{sec:yield_cuts}Yield cut}
\begin{figure}[!htb]
	\includegraphics[width=\columnwidth]{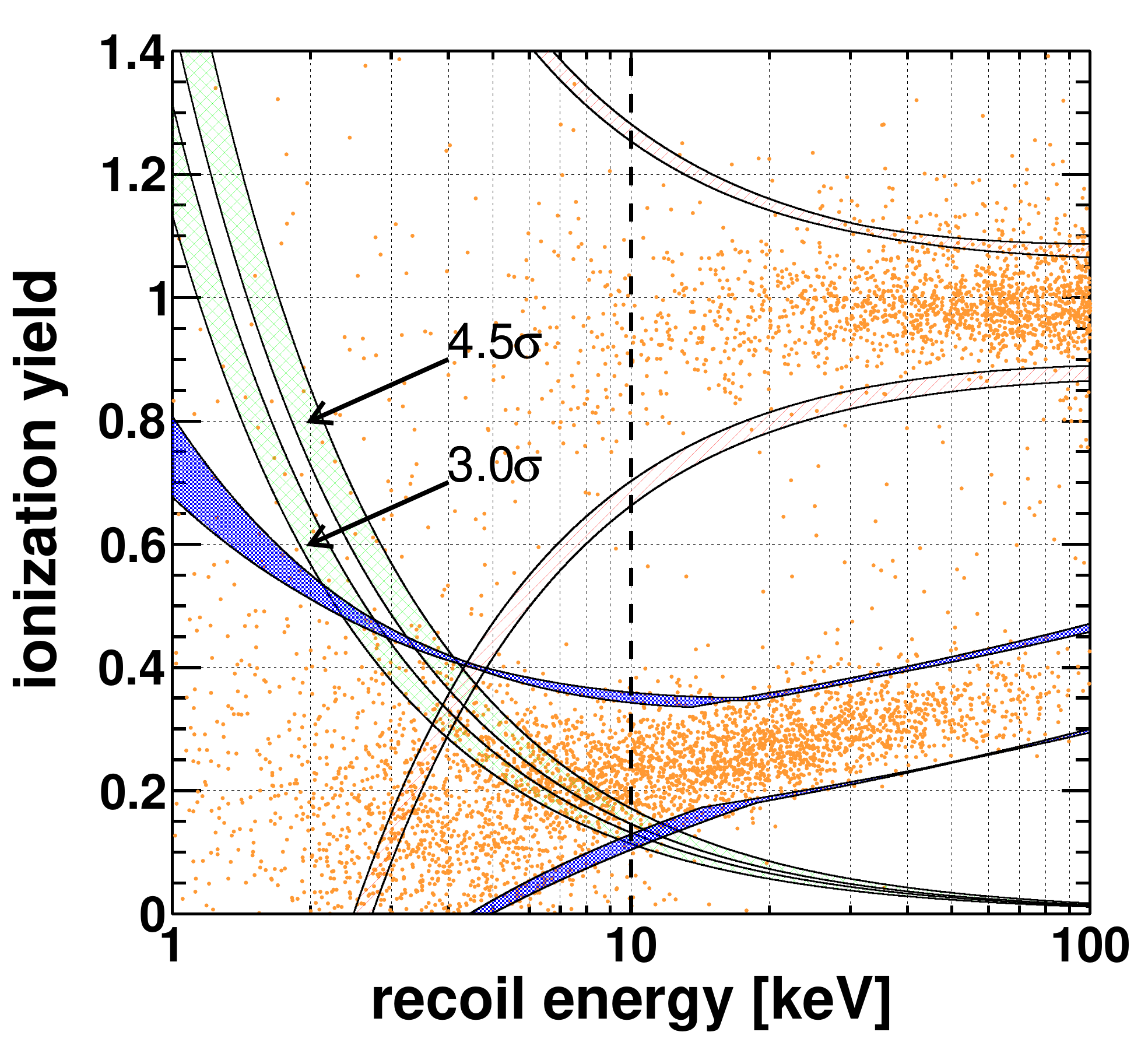}
	\caption{\label{fig:c58r_bands} (Color online) \er{} and \nr{} bands defined in the plane
	of yield versus recoil energy for a representative detector (T1Z2).  The \er{} band mean
	$\pm$3$\sigma$ curves (red hatched), \nr{} band mean $\pm$2$\sigma$ curves (blue fine
	cross-hatched), ionization threshold 3.0$\sigma$ and 4.5$\sigma$ curves (green coarse
	cross-hatched), and the 10\,\keVr{} threshold line (black dashed) are shown superimposed
	onto $^{252}$Cf data (points).  The widths of the bands represent the time-period
	variation of the fits that are used to define them.  
	}
\end{figure}

An energy-dependent cut on ionization yield is used as the primary method for discriminating
\nr{s} from \er{s}.  An \nr{} ``band'' was derived for each detector by fitting the distribution
of NR yields in $^{252}$Cf calibration with a Gaussian hypothesis in bins of recoil energy.  The
collections of best-fit Gaussian means and widths were then fit with energy-dependent functional
forms, giving smooth parametrizations versus recoil energy of the average yield and the yield
resolution for \nr{s}.  The functional form for the means was inspired by the Lindhard
theory~\cite{Lindhard63} ($y = a \cdot E_r^b$, where $a$ and $b$ are fitted parameters), whereas
the widths are fit to a power law below a fitted energy threshold and a constant above~\footnote{
Specifically the form $\sigma_y(E_r \leq \bar{E}) = c \cdot E_r^d$ and $\sigma_y(E_r > \bar{E}) =
e$ was used, where $c$, $d$, $e$, and $\bar{E}$ (the energy threshold) are fitted parameters.}.
\er{} bands were similarly constructed from $^{133}$Ba calibration.  The primary ionization-yield
cut requires that events be located within the $\pm$2$\sigma$ width of the \nr{} band.  By
construction, the selection efficiency is $\sim$95\% and roughly constant with energy (see
\fig~\ref{fig:c58r_combeff_chisq_58}).  Variations in detector response over the course of the
WIMP search caused slight variations in each detector's \nr{} and \er{} bands.  Consequently, the
bands depend on both detector and time.  The time variation is represented by widths of the band
lines shown in \fig~\ref{fig:c58r_bands}, where they are plotted with the $^{252}$Cf data from
which the \nr{} bands are derived.

To prevent \er{s} from entering the signal region at low recoil energy where the \er{} and \nr{}
bands overlap (see \fig~\ref{fig:c58r_bands}), the yield-based discrimination is refined to
include a requirement that candidate events lie at least 3$\sigma$ below the mean of the \er{}
band (see ~\sect~\ref{sec:data_defs}).  This condition and the ionization threshold have the
greatest impact on the overall \nr{} detection efficiency at low energies.  The efficiencies for
these two cuts are 100\% for recoil energies greater than $\sim$15--25\,\keVr{} (depending on
the detector) and decrease rapidly to zero for lower energies.  This behavior is illustrated in
\fig~\ref{fig:c58r_combeff_chisq_58}, where the efficiency of these two cuts is shown combined
with the other event-level quality cuts. 

\subsection{\label{sec:data_defs}Data regions}
To assess the effectiveness of the timing cuts defined in~\sect~\ref{sec:timing_analysis}, several
standard data regions in the nontiming variables were defined. This terminology will be used
throughout the rest of this work to describe data samples used for timing-cut tuning and
sideband-style background estimations.  All of the event samples include the data-quality and
fiducial-volume restrictions.  Further, they include the requirement that events have charge
energies greater than 4.5 standard deviations above the mean electronic-noise level.  
\begin{enumerate}
\item \textbf{Nuclear-recoil single scatter (NRSS).} Events that are below the \er{} band mean
-3$\sigma$ line, within the \nr{} band mean $\pm$2$\sigma$ lines, and are single scatters.  The
portion of the NRSS events that pass the timing cut and are in the range of 10--100\,\keVr{} make
up the WIMP signal region.  The \fivedchi{} timing cut uses a slightly modified signal region in
that the width of the \nr{} band is optimized.  We use the term NRSS there as well, leaving the
precise width of the \nr{} band to be determined from context. 

\item \textbf{Nuclear-recoil multiple scatter (NRMS).} Events that are below the \er{} band mean
-3$\sigma$ line, within the \nr{} band mean $\pm$2$\sigma$ lines, and are not single scatters.  

\item \textbf{Wide-band (WB).} Events that are below the \er{} band mean -5$\sigma$ line, and
above the \nr{} band mean +2$\sigma$ line.

\item \textbf{Wide-band multiple scatter (WBMS).} Events that are below the \er{} band mean
-5$\sigma$ line, above the \nr{} band mean +2$\sigma$ line, and are not single scatters.  These
events are a good representation of \seuh{s}.  
\end{enumerate}

\section{\label{sec:timing_analysis}Phonon Timing Discrimination}
ZIP detectors have an excellent ability to discriminate \nr{s} from \er{s} if the energy depositions
occur away from the surfaces of the detector.  But background \seuh{s} can populate the \nr{}
band, despite being \er{} in nature, since they can have reduced ionization yield.  We remove
\seuh{s} from our WIMP candidate sample using the timing characteristics of the phonon signals. A
small number of \seuh{s}, however, can survive into the signal region. We call these leakage
events.  The definition of the timing cut affects the \se{} leakage and the total exposure
(through the \nr{} acceptance), so a crucial element in the timing-cut construction is tuning for
optimal WIMP-detection sensitivity.  Three timing-cut constructions are reviewed in this section.

Having three independent methods for \se{} rejection gives a handle on the systematic uncertainty
of the leakage estimates.  The WIMP limits for these three realizations of \se{} rejection and
sensitivity maximization are presented in \sect~\ref{sec:results}. \tab~\ref{tab:variables} gives
a summary of the different timing parameters used for each timing cut construction: the
``classic,'' ``neural-network,'' and ``\fivedchi{}'' analyses.  The choice of parameters used in
particular timing-cut constructions is explained in the following sections. 
\begin{table}[!hbt]
\begin{tabular}{ c  c  c }
\hline
\hline
Quantity	& Description 	& Analysis	\\ \hline
$\tau$		& VF phonon rise time	   &  NN, \fivedchi{}	\\
$t_{\mathrm{del}}$	& VF phonon delay	   &  NN, \fivedchi{}	\\
$\tilde{\tau}$		& CF phonon rise time	   &  Classic, \fivedchi{} \\
$\tilde{t}_{\mathrm{del}}$	& CF phonon delay	   &  Classic, \fivedchi{} \\
$\tilde{\tau}_{\mathrm{4070}}$		& CF phonon 40--70\% rise time &  \fivedchi{}	\\
$\tilde{w}$	& CF phonon pulse width   &  NN	\\
$P_{\mathrm{5070}}$	& phonon 50--70\,kHz power   &  NN	\\
\hline
\hline
\end{tabular}
   \caption{\label{tab:variables}Brief description of each of the phonon timing quantities used in
   this work.  The abbreviation ``VF'' indicates the use of variable-frequency filtering prior to
   the application of the RT-FT-walk algorithm.  ``CF'' indicates a constant (50\,kHz) filter
   prior to that algorithm (see \sect~\ref{sec:par_extraction}).  Under ``Analysis,'' ``NN''
   refers to the neural-network timing analysis.
   }
\end{table}

Each of the timing cuts was optimized to produce the best expected sensitivity to WIMPs given the
expected leakage.  Since we do not know the WIMP mass, a ``target'' value is chosen for each
analysis (60\,\gev{} for most analyses in this work) and the expected sensitivity is maximized
given that WIMP mass.  The spectrum-averaged exposure (SAE) is a way to quantify the amount of the
raw exposure (MT = detector Mass $\times$ live Time) that is utilized toward the WIMP search over
the analysis energy range, given a WIMP recoil spectrum for mass $m_{\chi}$ of $f(E_r;m_{\chi})$.
The SAE is computed as follows:
\begin{equation}\label{E:sae}
\mathrm{SAE}(m_{\chi},E_l,E_h) = \mathrm{MT} \frac{\int_{E_l}^{E_h} dE_r \epsilon(E_r)f(E_r;m_{\chi})}{\int_{E_l}^{E_h} dE_r
f(E_r;m_{\chi})},
\end{equation}
where $E_r$ is the recoil energy, $E_l$ ($E_h$) is the lower (upper) signal-region energy limit,
and $\epsilon(E_r)$ is the cumulative signal acceptance efficiency for all cuts at a recoil energy
$E_r$.  Note that the SAE is equal to the raw exposure only when the analysis efficiency is unity
over the entire energy range and that only SAEs with the same WIMP-mass assumptions and
signal-region energy range are strictly comparable.  Sometimes the right-hand side of
\eq~(\ref{E:sae}) with MT divided out is called the spectrum-averaged efficiency.  The SAE is
computed on a detector-by-detector basis and then summed in the final analysis. 

Before looking at the events in the final signal region (see \sect~\ref{sec:unblinding} on
``unblinding'') we used the expected leakage and the SAE of each timing cut optimization to
calculate the expected sensitivity.  The expected sensitivity is computed by using the expected
leakage and calculating the upper limit of counts at the 90\% C.L. This is normalized by the SAE to
produce the lowest WIMP rate the experiment is sensitive to. For the 10\,\keVr{} threshold
analysis the \fivedchi{} timing cut had the best sensitivity, while our lower-threshold analysis
showed the classic method to have the best sensitivity.  The main results of this work are
therefore the limits of the \fivedchi{} analysis for WIMP masses above 11.3\,\gev{}, and the
limits of the classic method for WIMP masses below 11.3\,\gev{}. 
\subsection{\label{sec:classical_timing}Classic timing analysis}
The phonon timing cut strategy that was used in the original analysis of these, as well as earlier
CDMS data~\cite{Akerib2006, Ahmed2009, Ahmed2010a} was also used in our reanalysis and provides a
point of comparison between the two.  

\begin{figure}[!htb]
    \includegraphics[width=\columnwidth]{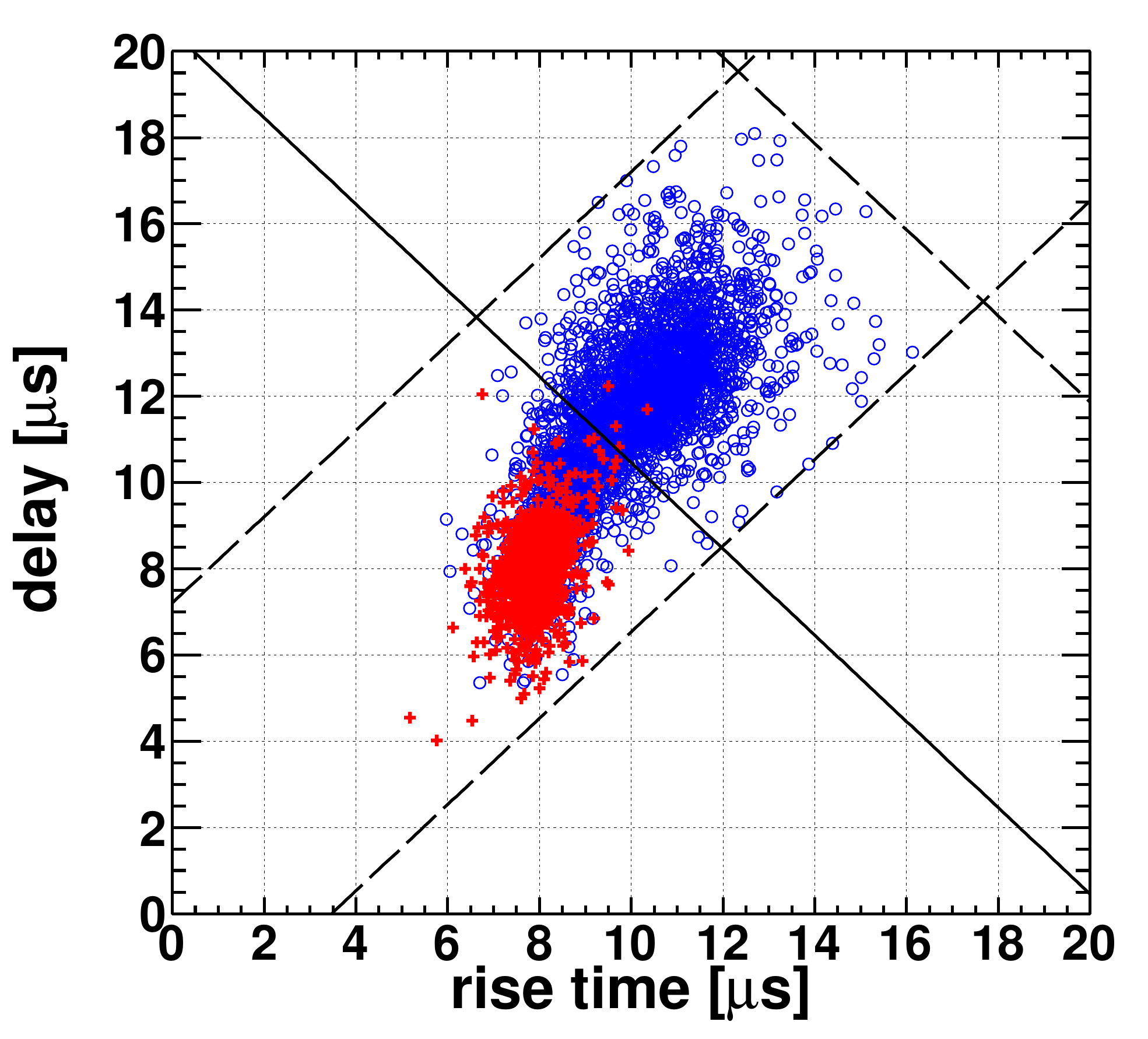}
    \caption{\label{fig:timing_2d} (Color online) Classic timing cut in the delay versus rise-time
    plane for a representative detector (T1Z5).  $^{252}$Cf NRSS and NRMS events (blue open
    circles) and $^{133}$Ba WBMS events (red crosses) are shown.  Accepted events lie within the
    ``consistency'' region (black dashed lines) and to the upper right of the discrimination cut
    (black solid line).  Considering the $^{133}$Ba WBMS events (red crosses), 2381 events fail
    the timing cut and four events pass. 
    }
\end{figure}

Our ``classic'' timing parameter is defined as the sum of two quantities: the delay
($t_{\mathrm{del}}$) and the 10--40\% rise time ($\tau$), both derived from the most energetic
phonon signal among the four sensors (see \sect~\ref{sec:reprocessing}).  The sum is approximately
the optimal combination of these two variables, as can be seen in \fig~\ref{fig:timing_2d}.  A
timing cut is defined as a set of detector-dependent thresholds on the distributions of this
parameter, below which all events are rejected.  The thresholds were determined by an optimization
scheme that approximately maximizes the sensitivity to a WIMP with a mass of 60\,\gev{}.
Practically, this was accomplished by maximizing the WIMP-search exposure (as measured with
$^{252}$Cf \nr{s}) while keeping the total leakage approximately equal to a ``target'' leakage of
$\sim$0.5 events.  This approximately maximized the 60\,\gev{} WIMP sensitivity while keeping the
total expected leakage well under one event~\cite{JianjieThesis}. 

\begin{figure}[!htb]
  \includegraphics[width=\columnwidth]{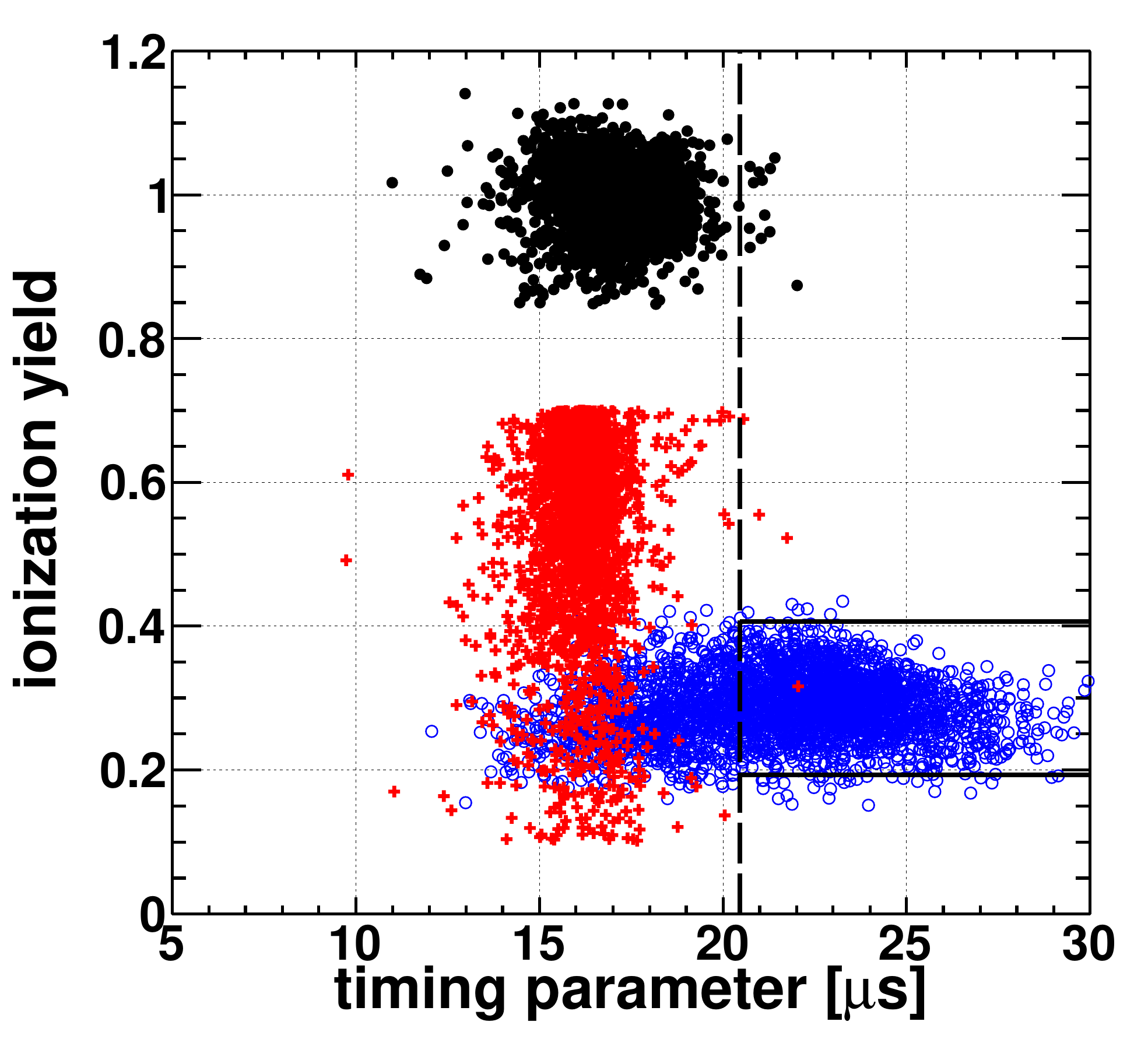}
  \caption{\label{fig:yield_timing} (Color online) Classic timing cut in the ionization-yield
  versus timing-parameter plane for a representative detector (T1Z5).  $^{252}$Cf NRSS and NRMS
  events (blue open circles), $^{133}$Ba WBMS events (red crosses), and $^{133}$Ba events in the
  \er{} $\pm$2$\sigma$ band (black filled circles) are shown.  Accepted events lie to the right of
  the timing parameter line (black dashed line) and in the \nr{} band (black solid lines).
  Considering the $^{133}$Ba WBMS events (red crosses), 2381 events fail the timing cut and four 
  events pass. 
  }
\end{figure}

The expected \se{} leakage was estimated from representative $^{133}$Ba calibration and sidebands
in the WIMP-search data that are insensitive to WIMPs.  The \se{} background estimates for the three 
timing-cut strategies are described in detail in \sect~\ref{sec:surface_event}.  Although the
quality cuts remove most unusual events, data-reconstruction artifacts occasionally result in
events with extreme kinematic quantities (``outliers'').  To prevent such outliers from skewing
the timing-cut optimization, a consistency cut rejects events for which
$\tau$\,+\,$t_{\mathrm{del}}$ is greater than 32\,$\upmu$s or $\tau$\,-\,$t_{\mathrm{del}}$ falls
outside the 0.5\% and 99.5\% quantiles of the combined $^{133}$Ba and $^{252}$Cf data sets.
Figure~\ref{fig:timing_2d} shows the classic timing cut in the delay versus rise-time plane, and
\fig~\ref{fig:yield_timing} shows the cut in the yield versus timing-parameter plane for an
example detector~\footnote{Note the \er{s} and \nr{s} have less clear ``surface discrimination''
within the respective populations.  The difference between \er{s} and \nr{s} arises
because Luke phonons produce a fast-rising component due to their ballistic nature and
consistent near-surface production component.  \er{s} give a greater fraction of Luke phonons, so
will generically be faster.}.  Applying this timing cut results in a total SAE summed over
detectors of 220\,\kgd{} between 10 and 100\,\keVr{} for a 60\,\gev{} WIMP and an expected
\se{} leakage of \nae{0.64}{0.17}{0.15} events, calculated after unblinding.  The post-unblinding
calculation is generally more accurate because it makes use of the previously sequestered (see
\sect~\ref{sec:unblinding}) \nr{} single scatters that failed the timing cut (see
\sect~\ref{sec:estimated_bknds}).

\subsection{\label{sec:nnet_timing}Neural-network timing analysis}
A neural-network technique was used to develop a timing cut using four timing parameters: the
previously defined phonon delay $t_{\mathrm{del}}$ (i) and rise time $\tau$ (ii); the phonon pulse
width $\tilde{w}$ (iii), defined as the time difference between the 80\% points on the rising and
falling edges of the largest-amplitude phonon pulse; and the spectral power of the largest phonon
pulse $P_{\mathrm{5070}}$ (iv), integrated between 50 and 70\,kHz~\cite{TommyThesis}.  These
parameters were chosen because they showed the most promising discrimination in their
one-dimensional distributions.

These four variables were fed into a principal component analysis~\cite{Pearson1901}.  Principal
component analysis is a statistical method for determining a unitary transformation that takes $N$
possibly correlated input vectors and returns $N$ output vectors that are linear combinations of
the input vectors ($N=4$ in this case).  The output vectors are ordered by their statistical
variance, so that the output vector with the $i{\textrm{th}}$-highest variance is called the
$i{\textrm{th}}$ principal component.  Since the input vectors can have different characteristic
scales and are dimensional, the input vectors were normalized to zero mean and unity variance so
that the ordering of statistical variances of the output vectors is meaningful in an absolute
sense.

Neural-network computational complexity scales poorly with the number of input parameters.
Therefore, only the first two principal components (i.e., those with the highest variance) were
selected as inputs for the neural network.  Given input parameters with similar intrinsic
resolution and physical relevance, the high-variance combinations will be those that maximally
separate the distinct populations (i.e., \nr{s} and \er{} \seuh{s}).  Principal components were
selected separately for each detector and for several bins of time spanning the WIMP-search data
set.  The latter is necessary to capture changes in detector performance caused by variations in
operating conditions. The principal component rotation showed that all four timing parameters
contribute significantly to the first and second principal components in most cases.   This
generally indicates that the use of the extra parameters (as compared to the classic analysis) is
beneficial even though in the end the sensitivity change is not dramatic (see
\sect~\ref{sec:indiv_limits}).

The neural network that was used is a multilayer perceptron with one hidden layer, 30 neurons,
and a logistic sigmoid activation function.  The NETLAB package~\cite{nabney} for
MATLAB was used to perform this analysis.  Training samples for \seuh{s} and \nr{s} were
selected from the $^{133}$Ba and $^{252}$Cf calibration data, respectively.  Events from the \nr{}
training sample were assigned a target output value of 1, while \seuh{s} were assigned a target
value of 0.  Data were sorted into two bins of recoil energy, above and below 30\,\keVr{}, in order
to take the energy dependence of the timing parameters into account.  Finer energy binning was not
possible because of small statistics in the training samples.

\begin{figure}[!htb]
   \includegraphics[width=\columnwidth]{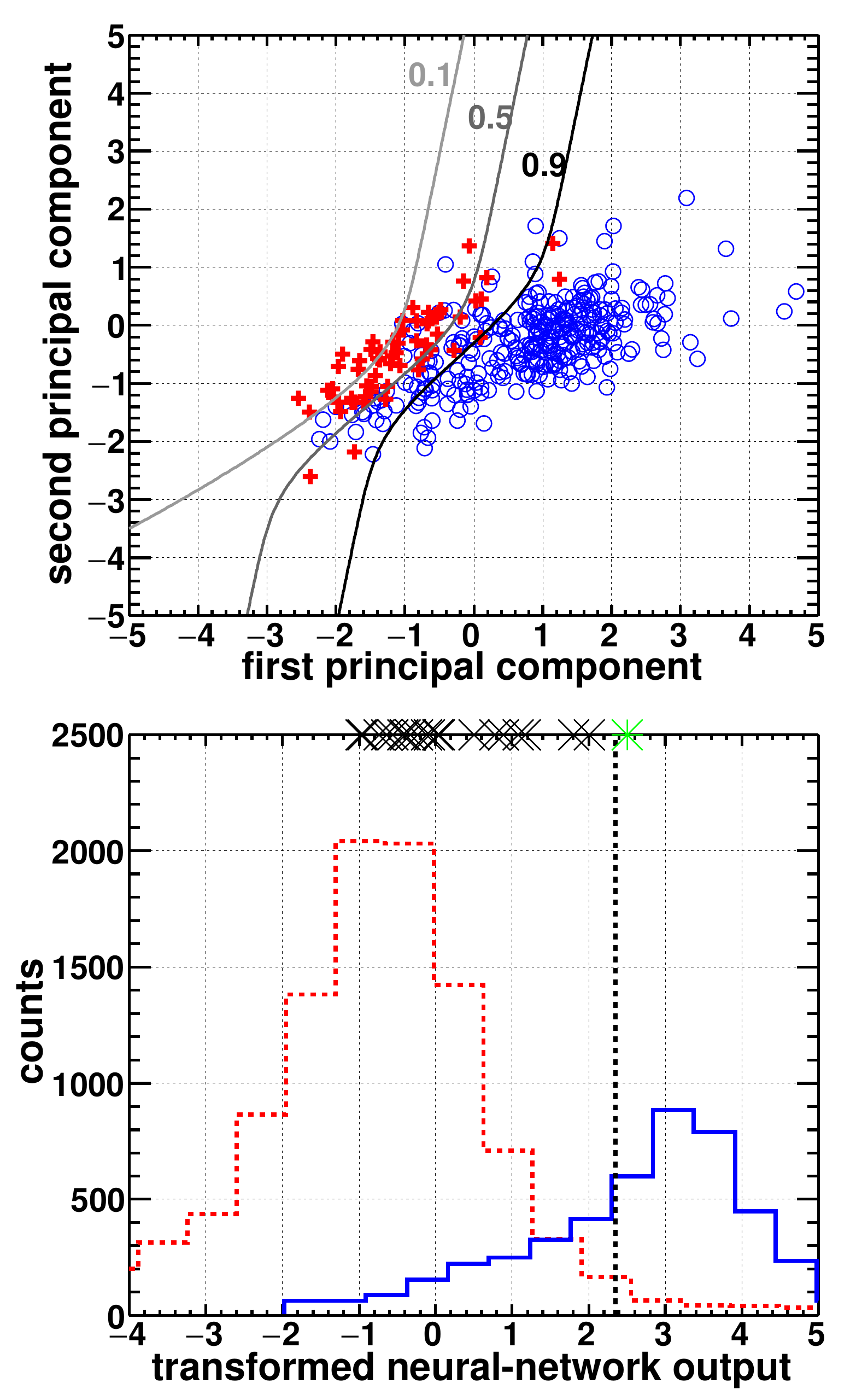}
   \caption{\label{fig:nnet_dist_405} (Color online) Neural-network training distributions of a
   representative detector (T1Z5), for the low-energy neural-network bin ($<$\,30\,\keVr{}). (Top
   panel) Contours of constant neural-network output (grayscale solid curves, see text) in the
   plane of the first two principal components.  $^{252}$Cf NRSS and NRMS events (blue open
   circles) and $^{133}$Ba WBMS events (red crosses) are also shown.  (Bottom panel) Distributions
   of the ``transformed'' neural-network output: the distribution of \nr{s} from the $^{252}$Cf
   data mentioned above (blue solid); and the \se{} distribution from the $^{133}$Ba data
   mentioned above (red dashed).  The cut threshold (vertical black dashed line) with the
   WIMP-search data passing (green asterisk) and failing (black $\times$'s) the timing cut are
   also plotted along the top of the plot.  The transformed neural network is used to make the
   distribution separation more visible; it is computed by using the inverse of the neuron
   response function: $\phi:(-\infty,\infty)\rightarrow(0,1)$ with $\phi(x) = 1/(1+e^{-x})$
   }
\end{figure}

Separate neural networks were trained for each combination of detector, time period, and energy
bin using the standard back-propagation-of-errors training algorithm~\cite{nabney}.  Once the
neural networks were trained, they assigned a numerical value to each event in the range 0--1.  An
output value close to 1 (0) corresponds to events of \nr{}-like (\se{}-like) character.  The
distributions of this output parameter for the calibration data, as seen in
\fig~\ref{fig:nnet_dist_405}, were then used to set a threshold for each detector, time period,
and energy bin.

The thresholds were set such that the WIMP-search exposure is maximized using a target \se{}
leakage of 0.5 events. As in the classic analysis, this target leakage approximately maximized the
60\,\gev{} WIMP sensitivity and the same optimization procedure was used.  Following optimization,
the total SAE for a 60\,\gev{} WIMP is about 216\,\kgd{}, with an expected leakage of
\nae{0.87}{0.24}{0.21} events.  Similar to the leakage estimate for the classic timing cut, this
expected leakage was estimated after unblinding for better accuracy (see
\sect~\ref{sec:estimated_bknds}).

\subsection{\label{sec:5dchi_timing}\fivedchi{} timing analysis}
The \fivedchi{} \se{} rejection~\cite{JosephThesis} was implemented by differentiating events
based on a goodness of fit to two event-type hypotheses~\cite{JeffThesis,JoelThesis}:
$\chi^2_{\mathrm{N}}$ for \nr{}, and $\chi^2_{\mathrm{B}}$ for \seuh{s}.  Five timing quantities
were used to form each $\chi^2$ value.  Three of the quantities are measures of rise time for the
largest-amplitude phonon channel: two based on the 10--40\% rise time ($\tau$ and $\tilde{\tau}$);
and one based on the 40--70\% rise time ($\tilde{\tau}_{\mathrm{4070}}$).  The remaining two
quantities correspond to the delay of the phonon pulse relative to the prompt charge pulse -- one
computed using a variable and one a constant-frequency RT-FT walk ($t_{\mathrm{del}}$ and
$\tilde{t}_{\mathrm{del}}$).  The inclusion of different measures of the same physical quantities
(delay and rise time here) increases the robustness of the $\chi^2$ value and is more effective at
identifying outliers.  Therefore, the delay and rise-time parameters with good one-dimensional
discrimination and the least redundancy (correlation) were chosen. 

Event samples of neutrons and \seuh{s} taken from calibration data were used to constrain the
timing-quantity distributions for each event type.  For each detector, calibration data were 
separated into neutron, charge-side \se{}, and phonon-side \se{} samples.  The energy-dependent
means, $\bm{\mu}(E_r;\alpha)$--a vector of the timing-quantity distribution means for each event
type $\alpha$--were then fit to the empirically motivated functional form
\begin{equation}\label{chieq1}
\bm{\mu}(E_r;\alpha) = \bm{a}_{1}(\alpha) + \bm{a}_{2}(\alpha) E_r^2 + \bm{a}_{3}(\alpha) \sqrt{E_r}, 
\end{equation}
where the $\bm{a}_{i}(\alpha)$ are free parameters for each timing quantity (the vector indices)
and for particle type $\alpha$.  The $\sqrt{E_r}$ term is observed to improve the fit.  The
covariance matrix $\bm{\sigma}(E_r;\alpha)$ was similarly fit using

\begin{equation}\label{chieq2}
\bm{\sigma}^2(E_r;\alpha) = \bm{b}_{1}(\alpha) + \frac{\bm{b}_{2}(\alpha)}{E_r^2}, 
\end{equation}
where the $\bm{b}_{i}(\alpha)$ are matrices of the free parameters for each pair of timing
quantities and particle type $\alpha$.  The functional form of the variance was motivated by
noting that in a simple model of a pulse with a linear rise but constant rise time, the
probability for a noise fluctuation before the pulse rises above the noise is inversely
proportional to the slope, and therefore is proportional to the inverse of the amplitude (energy).

The $\chi^2_{\alpha}$ was then formed for every event according to the following formula:
\begin{equation}\label{chieq3}
    \chi^2_{\alpha}(E_r) = (\bm{\xi}-\bm{\mu})^T\cdot (\bm{\sigma}^2)^{-1}\cdot (\bm{\xi}-\bm{\mu}).
\end{equation}
Here $\bm{\xi}$ is the vector embedding the five timing variables for each event and the
dependence on $E_r$ and $\alpha$ has been left implicit on the right-hand side.

The \se{} goodness-of-fit variable $\chi^2_{\mathrm{B}}$ was constructed for each event by the
definition
\begin{equation}
  \chi^2_{\mathrm{B}} \equiv \text{min}(\chi^2_p,\chi^2_q), 
\end{equation}
where $\chi^2_q$ and $\chi^2_p$ are the charge-side \se{} and phonon-side \se{} goodness-of-fit
variables respectively.

\begin{figure}[!htb]
\centering
   \includegraphics[width=\columnwidth]{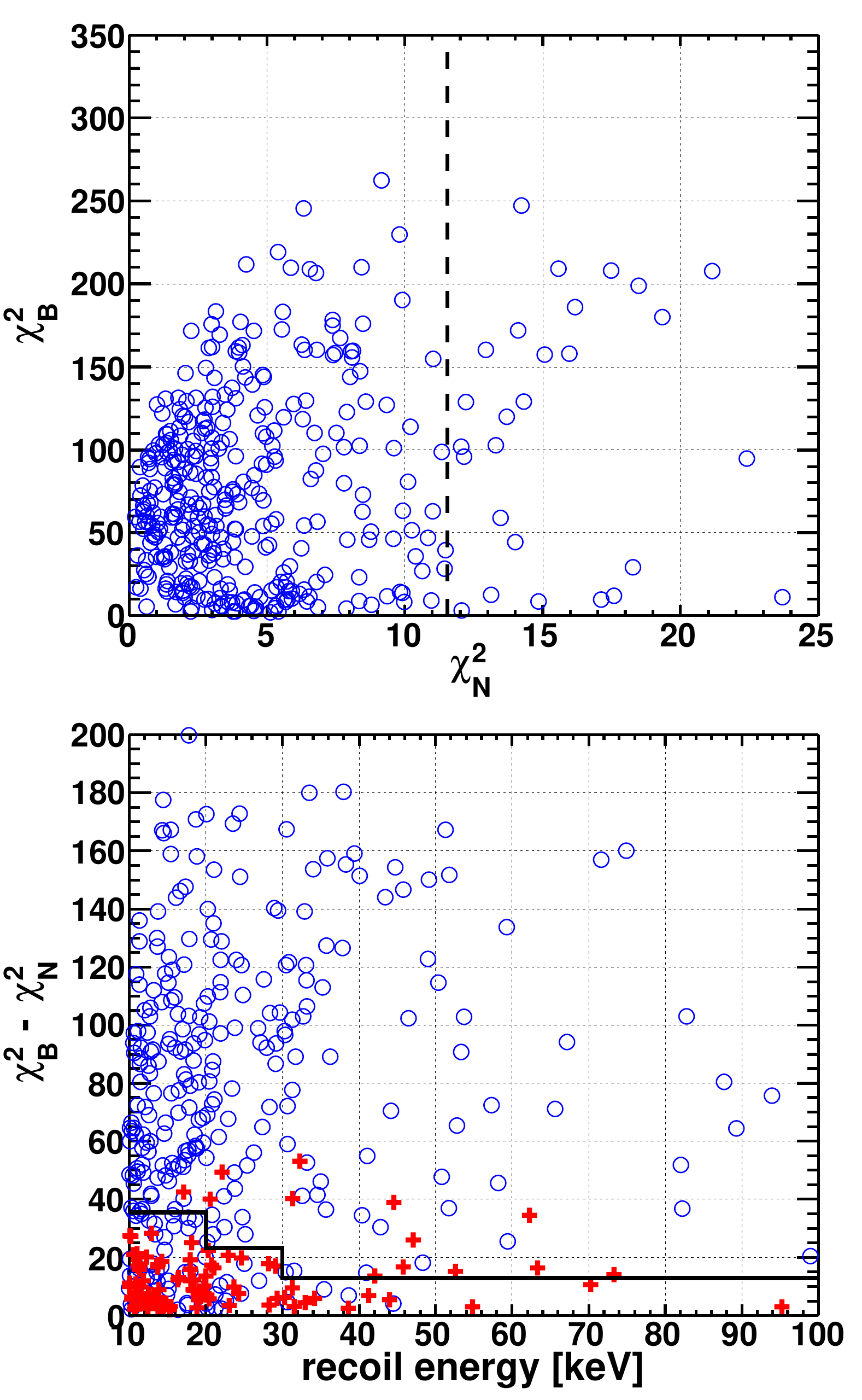}
   \caption{\label{fig:chi2bchi2n_and_diff} (Color online) Distributions of calibration data in
   the \fivedchi{} parameters for a representative detector (T1Z2). (Top panel) $^{252}$Cf events
   in the \nr{} $\pm$2$\sigma$ band (blue open circles) are shown with the consistency cut that
   retains 90\% of neutrons (vertical black dashed line) in the $\chi^2_{\mathrm{B}}$ versus
   $\chi^2_{\mathrm{N}}$ plane.  (Bottom panel) Events passing the consistency cut--\nr{s} from
   $^{252}$Cf data (blue open circles) and \seuh{s} from $^{133}$Ba data (red crosses)--in the
   $\chi^2_{\mathrm{B}}-\chi^2_{\mathrm{N}}$ versus recoil-energy plane.  The energy-dependent
   \fivedchi{} timing cut is also shown (black solid line) and events above the line pass the cut.
   Note the cut is tighter at low energies and looser at higher energies.  
   }
\end{figure}

Two restrictions were set in the plane of $\chi^2_{\mathrm{B}}$ versus $\chi^2_{\mathrm{N}}$ that
together complete the definition of the \fivedchi \se{} rejection cut (see
\fig~\ref{fig:chi2bchi2n_and_diff}). First the potential WIMP events were required to have
$\chi^2_{\mathrm{N}} \leq c_i$, where $c_i$ is a value for the $i{\textrm{th}}$ detector, set by
requiring that 90\% of the calibration-data neutrons pass the cut. To distinguish potential signal
events from \seuh{s}, it was required that $\chi^2_{\mathrm{B}}-\chi^2_{\mathrm{N}}\geq \eta_i(e)$
where the index $i$ indicates the detector number and the parameter $e$ indicates the event's
energy bin (10--20\,\keVr{}, 20--30\,\keVr{}, or 30--100\,\keVr{}). 

We parametrize--using adjustable parameters $t$ and $b$--the ionization-yield ($y$)
restriction as 
\begin{equation}
\mu_{nr}-b \sigma_{nr} \leq y \leq
\mu_{nr}+t\sigma_{nr},
\end{equation}
where $\mu_{nr}(E_r)$ and $\sigma_{nr}(E_r)$ are the energy-dependent mean and standard deviation
of the ionization yield for \nr{s}, as found using neutron calibration data.  The parameters $t$ and
$b$ were required to be the same for all detectors and energy bins.  The values $\eta_i(e)$, $t$,
and $b$ were determined by a simultaneous optimization for all detectors and energy bins that
maximizes the total SAE for 60\,\gev{} WIMPs. 

Optimization of the \fivedchi{} timing cut and the \nr{} band definition was based on requiring
the best overall expected sensitivity.  For a given timing cut (the set $\{\eta_i\}$) and \nr{}
band definition, the expected sensitivity is constructed by dividing the 90\% Poisson upper limit
on the expected leakage by the SAE.  For an individual detector we denote the leakage as
$\mathcal{L}_i(\eta_i)$ and the SAE as $\mathcal{S}_i(\eta_i)$.  These are smooth functions
computed by fitting leakage and exposure evaluations at discrete $\eta_i$ using  $^{133}$Ba
calibration data. 

Because the cut is defined such that decreasing $\eta_i$ will loosen the restriction, both the
leakage and the SAE for each detector are monotonically decreasing--leading to the condition
that the slopes $d\mathcal{S}_i/d\mathcal{L}_i$ for the optimum cut are equal.  If not, a unit
increase in the leakage of a detector with larger slope could be offset by a unit decrease in a
detector with a smaller slope with a net increase in exposure, improving the sensitivity.  The
timing cut optimization was done with respect to this slope, which parametrizes the $\{\eta_i\}$
uniquely.

For the optimum set of timing parameters $\{\eta_i\}$ the \nr{} band definition is selected by
choosing the values of $t$ and $b$ that optimize the overall expected sensitivity.

The yield and timing cuts that optimize the expected sensitivity for a 60\,\gev{} WIMP produced an
asymmetric \nr{} band cut with $b=-1.9$ and $t=1.8$.  The sensitivity optimization gives a total
expected leakage of 0.5 events and a total SAE of 250\,\kgd{} given a WIMP mass of 60\,\gev{}.
The estimated leakage after unblinding is \nae{1.19}{0.23}{0.21} events.  

\subsection{\label{sec:timing_ex}Extended analyses}
There has been growing interest in low-mass WIMP searches because of some intriguing published
results~\cite{Agnese2013,Aalseth2013} and the suggestion that the baryon asymmetry is reflected in
the dark matter sector~\cite{Zurek2009}.  While many previous WIMP searches--guided by the SUSY
neutralino parameter space--paid much attention to $\sim$100\,\gev{} WIMP masses, it is
interesting in light of these new results to examine data with techniques optimized for much lower
WIMP masses $\sim$10\,\gev{}.  For this reason the timing-cut constructions considered so far were
extended with lower thresholds (down to 5\,\keVr{} for some detectors), in order to improve
sensitivity to low-mass WIMPs.

All of the timing cuts presented so far were optimized for a 60\,\gev{} WIMP mass, and the
corresponding analyses were restricted to a recoil-energy threshold of 10\,\keVr{}.  For events
above this threshold the best estimates of \se{} background leakage are about one event.  While
the \se{} leakage is worse for the extended-threshold analyses, for low WIMP masses sensitivity
improves considerably because of the steeply rising WIMP spectrum.  Simple extensions to the three
timing analyses were accomplished by lowering the thresholds, confirming the cut efficiencies
below 10\,\keVr{}, and reevaluating the \se{} leakage estimates.  The analysis region for the
extended analyses is approximately 5--15\,\keVr{} (threshold differs by detector; see below),
since recoils of light WIMPs ($\lesssim$10\,\gev{}) with energies greater than 15\,\keVr{} are
very rare. 

For the extended analyses there are several small changes to the event selections that help to
maximize the sensitivity to WIMP masses below 10\,\gev{}.  

A careful study of the leakage induced by lowering the charge threshold showed that for light
WIMPs with similar cross sections to the \cdmstwo{} silicon result~\cite{Agnese2013} a net gain is
obtained by lowering the charge threshold to 3.0 standard deviations above the mean noise value.
We therefore modified the charge threshold to this value for the extended analyses.  The
recoil-energy threshold was then effectively set by the condition that signal events have an
ionization yield below the mean -3$\sigma$ \er{} band line.  Of course, signal events were still
required to be within the mean $\pm$2$\sigma$ \nr{} band (with the usual slight modification for
the \fivedchi{} analyses).  Finally, the signal regions were taken to range from threshold up to
15\,\keVr{} since this is the region that is expected to add significantly to the low-WIMP-mass
exposure.  These changes are implicit where we use the abbreviation ``NRSS'' in the context of the
extended analyses.  Most of the timing cuts (except the \fivedchi{}, see below) did not have their
decision boundaries reoptimized for lower WIMP masses; the thresholds were simply extended and the
changes above were made.  

The \fivedchi{} method, in addition to lowering the thresholds, was partially reoptimized for a
8\,\gev{} WIMP mass.  The definitions of the $\chi^2$ variables
($\chi^2_\mathrm{N}$,$\chi^2_\mathrm{B}$) remained the same.  Additionally, the means and
covariances were taken to have the same functional dependences given in \peq~(\ref{chieq1})
and~(\ref{chieq2}).  The optimized yield cut was an asymmetric cut, with parameters $t=1.8$ and
$b=-1.9$, the same as the regular \fivedchi{} analysis.  The reoptimized sensitivity for the
8\,\gev{} was not significantly better than the standard \fivedchi{} analysis, and not as good as
the sensitivity for the classic extended-threshold analysis.  Therefore, this reoptimization was
not carried any further and the 60\,\gev{} optimized version is used.  

\subsection{\label{sec:final_timing_efficiency}Timing efficiencies and summary} 
\begin{figure}[!htb]
	\includegraphics[width=\columnwidth]{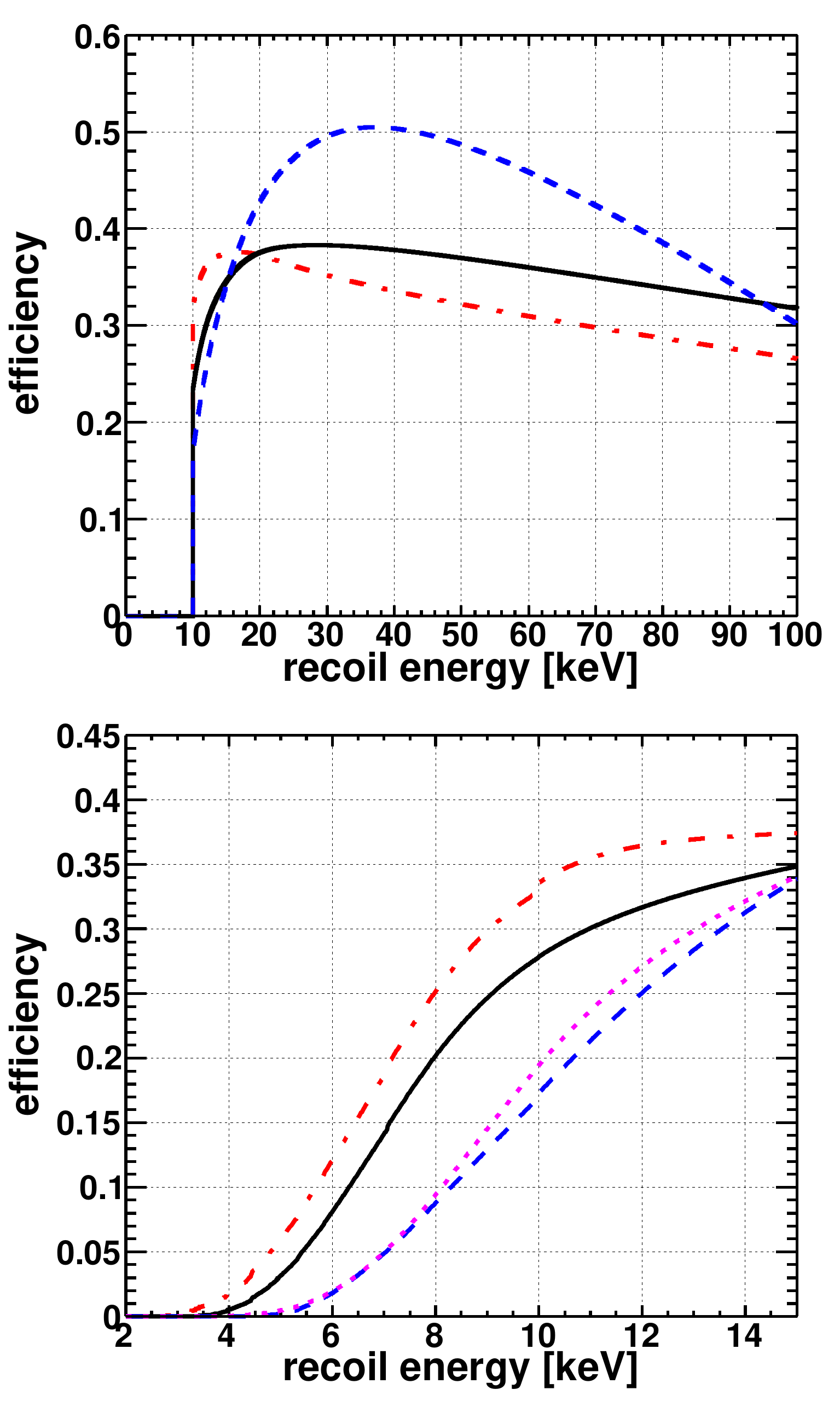}
	\caption{\label{fig:timing_effs}(Color online) Total combined efficiencies for all timing
	analyses.  (Top panel) Efficiencies for the 10\,\keVr{} threshold analyses.  The SAEs
	(60\,\gev{} WIMP) are 219.1\,\kgd{} for the classic analysis (black solid), 216.4\,\kgd{}
	for the neural-network analysis (red dot-dashed), and 262.3\,\kgd{} for the \fivedchi{}
	analysis (blue dashed). (Bottom panel) The efficiencies for the extended-threshold
	analyses with the addition of the \fivedchi{} analysis optimized for a 8\,\gev{} WIMP mass
	(magenta dotted). All of the analyses have a total exposure (before efficiency reductions)
	of 612.2\,\kgd{}. 
	}
\end{figure}

Energy-dependent efficiencies for the timing analyses were computed using neutrons from the
${}^{252}$Cf calibration data.  Figure~\ref{fig:timing_effs} presents the combined efficiencies of
all cuts (quality, fiducial volume, yield, and timing) for each timing analysis.  Above a
10\,\keVr{} threshold, the \fivedchi{} analysis has the best sensitivity and the highest SAE for a
60\,\gev{} WIMP (see \fig~\ref{fig:timing_effs} caption).  The classic analysis has the
second-best SAE, and provides continuity with the original analysis of this data
set~\cite{Ahmed2010a}.  The neural-network analysis yields the smallest SAE, but has the best
efficiency just above a 10\,\keVr{} recoil energy threshold.  For the extension of the analysis to
below 10\,\keVr{}, the classic timing analysis has the best sensitivity (though not the highest
efficiency at low recoil energies; see \fig~\ref{fig:timing_effs}).  As mentioned in
\sect~\ref{sec:timing_analysis}, the \fivedchi{} is our ``primary'' method above 10\,\keVr{} and
the classic is the primary extended-threshold method. 

The efficiency function is a necessary ingredient for producing the limits and any uncertainty on
this function is also present in the final limit.  The trigger efficiency uncertainty is $\sim$1\%
across the energy range; this is mostly statistical uncertainty.  Our quality cut efficiency is
calculated based on baseline noise levels using large event populations and so has negligible
uncertainty.  Efficiencies of the other cuts are measured by selecting neutron populations in
$^{252}$Cf data and observing the decrease in the population by the application of the cuts in
bins of recoil energy.  The results are then fit with an empirical functional form with a
low-energy falloff similar to the error function.  Fiducial-volume, ionization-yield and
phonon-timing cuts each contribute about a 5\% statistical uncertainty.  This measurement method
is, however, prone to error due to the fact that neutrons can have multiple scatters inside a
detector, while WIMPs cannot.  In the case of the fiducial-volume cuts we used a Monte Carlo
simulation to find a 5\% discrepancy for multiple scatters, and corrected for it.  Overall we
assign a conservative 3\% systematic uncertainty on the fiducial-volume and yield cuts for the
effect of multiple scattering.   Based on these estimates, using the averaged sizes of each
efficiency, we expect the total uncertainty on the efficiencies to be $\sim$6\%.  Generally this
is negligible on the scale that our final limits are presented.  A specific study of the
efficiencies for the extended-threshold analysis below 6\,\keVr{} showed that for the
extended-threshold limits this total uncertainty becomes $\sim$10\% at  the lowest WIMP mass
(6.26\,\gev{}) and drops to $\sim$7\% by a WIMP mass of 7\,\gev{}.

\section{\label{sec:unblinding}Unblinding}
A blinding technique was used to avoid bias in the setting of data selection cuts.  The current
work used what can be referred to as ``hidden signal box'' analysis~\cite{Klein2005} which is
common in rare-event searches where the signal region is known \textit{a priori}.  The same data
were analyzed previously~\cite{Ahmed2010a}, but since all the data were reprocessed with an
upgraded charge reconstruction algorithm, and the cuts were optimized solely based on calibration
data and distributions outside the newly masked signal region, this is a good approximation of a
hidden signal box analysis.  Signal events were hidden by removing the single-scatter events in
the \nr{} yield band.  This technique is effective for removing bias in the timing-cut
preparation, but restricts the information that can be used for estimating \se{} leakage before
unblinding has occurred.  For this reason the leakage estimates were done before and after the
unblinding, but consistency between the two methods is checked.  Including the post-unblinding
version using the \nr{} singles that fail the timing cut made the final estimate more robust.

\begin{figure}[!htb]
\includegraphics[width=\columnwidth]{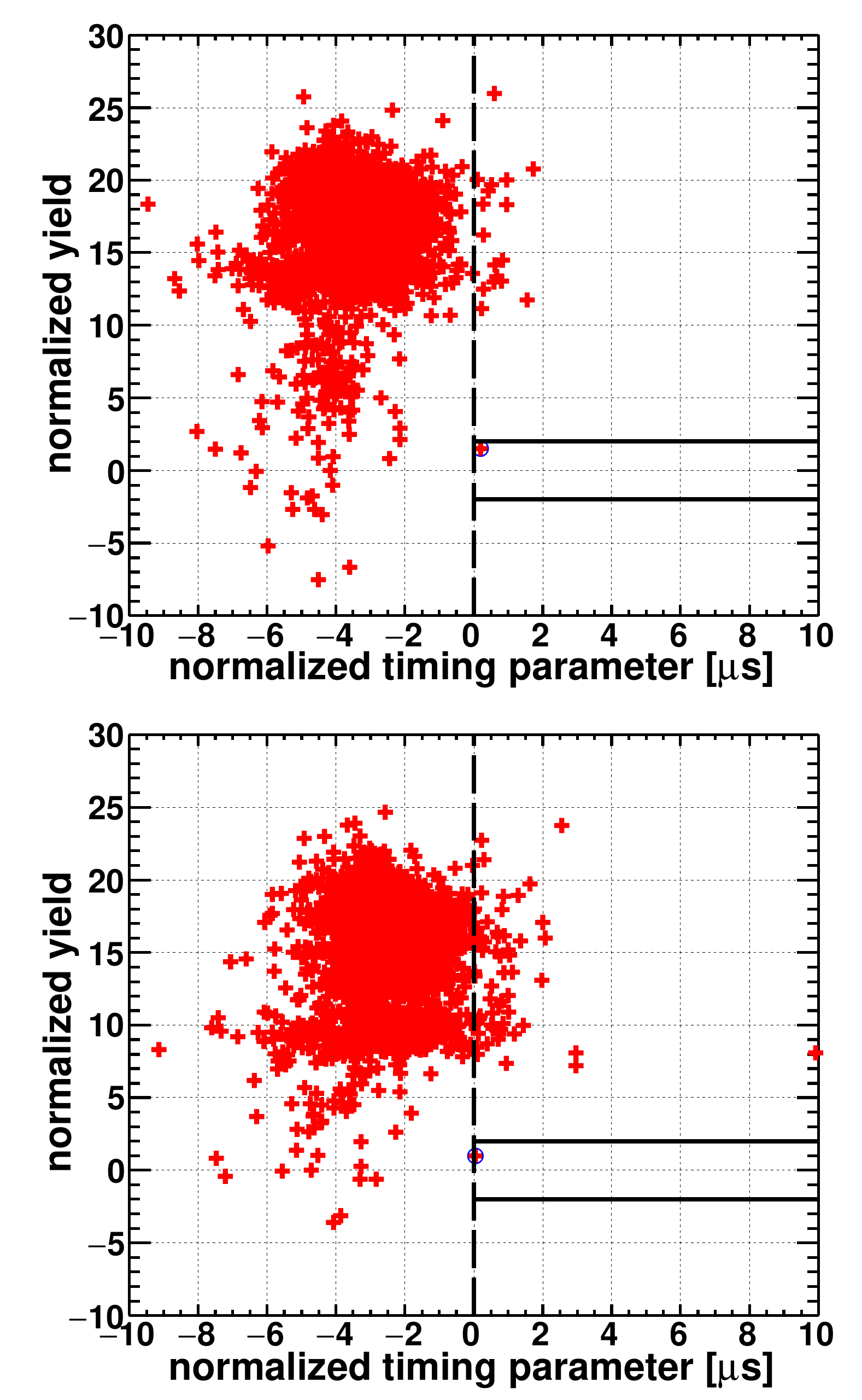}
   \caption{\label{fig:c58r_ws_yield_timing_classic_405_413}(Color online) 
   Distributions of WIMP-search events in the detectors containing the two candidate events for
   the classic timing analysis: T1Z5 (top panel) and T2Z3 (bottom panel).  The normalized yield is
   the distance from the average \nr{} yield in units of the standard deviation of the yield
   distribution; the black horizontal lines on the right side of the plot indicate the
   $\pm$2$\sigma$ \nr{} band.  The normalized timing parameter is the standard timing parameter
   ($t_{\mathrm{del}}$+$\tau$) minus the value of the cut boundary.  WIMP-search events (red
   crosses) to the left of the timing cut (vertical black dashed line) pass all cuts except phonon
   timing and yield; events to the right of that line pass the timing cut, and those that are also
   in the \nr{} band (solid black lines) pass the yield cut.  The highlighted event (blue open
   circle surrounding red cross) in the top panel is candidate 1 and the highlighted event in the
   bottom panel is candidate 2 (see \tab~\ref{tab:candidates}).  All the events shown are in the
   recoil energy range 10--100\,\keVr{}.  
   }
\end{figure}

Upon unblinding, the following number of events pass all cuts above 10\,\keVr{}: zero for the
\fivedchi{} timing analysis; two for the classic timing analysis; and one for the neural-network
analysis.  One of the two candidates from the original analysis is a candidate in both the classic
timing and the neural-network analyses~\cite{Ahmed2010a}.  The second candidate from that
analysis, whose poor charge-pulse fitting prompted this reanalysis (see
\sect~\ref{sec:reprocessing}), failed all timing analyses by a substantial margin.  Information
about the two passing candidate events is shown in \tab~\ref{tab:candidates}.
Figure~\ref{fig:c58r_ws_yield_timing_classic_405_413} shows the location of these events within
the signal region of the classic analysis. 
\begin{table}[!hbt]
\begin{tabular}{ c  c  c  c  c }
\hline
\hline
Detector	& Recoil energy [\keVr{}] & Yield		& Analysis		\\ \hline
T1Z5		& 12.30				   &  0.33		& NN, Classic, 2010	\\
T2Z3		& 10.81				   &  0.33		& Classic		\\ 
T3Z4		& 15.35				   &  0.26		& 2010  		\\ 
\hline
\hline
\end{tabular}
   \caption{\label{tab:candidates}Information about the three WIMP candidate
   events above 10\,\keVr{}.  Under ``Analysis,'' ``NN'' refers to the
   neural-network timing analysis and ``2010'' refers to the original analysis
   of these data~\cite{Ahmed2010a}.  Note that no candidate events are observed
   for the \fivedchi{} analysis.  
   }
\end{table}

Our extended-threshold analyses gave a wide range in terms of the number of candidate events.  The
classic analysis had six candidates and the neural-network analysis had 16.  No events were observed
in the extended version of the \fivedchi{} analysis.  A description of the candidate events for
our primary extended-threshold analysis (the ``classic'' cut) is given in
\tab~\ref{tab:mt_candidates}. The large number of candidate events in the neural-network analysis
is attributed to an increased leakage of anomalously low-ionization events, those that are
normally below the ionization threshold in the 10\,\keVr{} analysis~\cite{TommyThesis}. This
increased leakage of essentially zero-ionization events was not expected and is presumably due to
a bias in the training set of the neural network.  It has negligible contribution above the
original 4.5 standard deviation charge threshold.  The low number of events in the \fivedchi{}
analysis is not unexpected because that cut has an independent energy bin at 10--20\,\keVr{},
where the cut is rather stringent because of increasing leakage at low energy and the optimization
to exposure at a 60\,\gev{} WIMP mass. 
\begin{table}[!hbt]
\begin{tabular}{ c  c }
\hline
\hline
Detector	& Recoil energies [\keVr{}] \\ \hline
T1Z5		& 3.45, 5.73, 12.30	\\
T2Z3		& 10.81	\\ 
T4Z4		& 7.56	\\ 
T4Z5		& 7.25	\\ 
\hline
\hline
\end{tabular}
   \caption{\label{tab:mt_candidates}Information about the six WIMP candidate events for the
   ``classic'' extended-threshold analysis.  
   }
\end{table}

The post-unblinding leakage estimates can be found in \tab~\ref{tab:allleakage} along with the
analysis energy ranges, candidate numbers, exposures and WIMP-mass optimization assumptions.
{\renewcommand{\arraystretch}{1.2}
\begin{table*}[!hbt]
\begin{tabular}{ c  c   c  c  c  c  c }
\hline
\hline
Method	& Energy range [\keVr] & Exp. leakage & Candidates	&  WIMP mass [\gev{}]
& SAE [\kgd{}]  \\ \hline
Classic		&10--100	& 0.64$^{+0.17}_{-0.15}$ & 	2		& 60		& 220	\\
Neural Network	&10--100	& 0.87$^{+0.24}_{-0.21}$ &	1		& 60		& 216	\\
\fivedchi	&10--100	& 1.19$^{+0.23}_{-0.21}$ & 	0		& 60		& 250	\\
Classic ext.	&5--15	& $>$~1.48$^{+0.20}_{-0.20}$ & 	6		& 60		& 186	\\
Neural Network ext.	&5--15	& $>$~1.39$^{+0.21}_{-0.21}$ & 	16		& 60		& 190	\\
\fivedchiex	&5--15	& $>$~0.97$^{+0.15}_{-0.15}$ & 	0		& 60		& 202	\\
\fivedchiex	&5--15	& $>$~1.82$^{+0.31}_{-0.31}$ & 	0		& 8		&
3.78\footnote{SAE depends on an assumed WIMP mass [see \eq~(\ref{E:sae})]; we use the optimization
mass in all cases.  For this reason SAE is only comparable in situations of common optimization
mass and signal-region energy range}	\\ 
\hline
\hline
\end{tabular}
   \caption{\label{tab:allleakage}Expected leakage and exposure statistics for all of the \se{}
   rejection methods described in this work.  For the extended-threshold analyses, we quote
   5\,\keVr{} as an approximate lower limit on the signal region.  The actual threshold depends on the
   detector and is set by the crossing of the \er{} band limits and the charge threshold curves
   (see \sect~\ref{sec:timing_ex}).  The symbol $>$ is used to indicate lower limits on the
   expected leakage.  In those situations the event sets that are typically used to estimate the
   leakage do not have events all the way down to the signal-region threshold (see text). 
   }
\end{table*}
}

\section{\label{sec:estimated_bknds}Estimated Backgrounds}
\subsection{\label{sec:surface_event}Surface electron-recoil background}
Surface electron recoils originate from several sources: 1) particles emitted from $\beta$
emitters contaminating the surfaces of the detector and the material around it (notably
\ce{^{210}Pb}), 2) photo-electrons emitted from material neighboring the detector
through Compton scattering or the photoelectric effect, and 3) photons that interact in the
detector within a few microns of the surface. Photons from category 3) can be low-energy xrays
or high-energy photons that Compton scatter in the detector.  Past studies showed that the
dominant contributions are from \ce{^{210}Pb} and photon-induced backgrounds, which contribute
approximately equally. No other sources were found to be statistically significant. 

The expected number of \seuh{s} leaking into the WIMP signal region was calculated using the
number of single scatters in the \nr{} band that are rejected by the timing cut and the \se{}
rejection efficiency.  The \se{} rejection efficiency was estimated from three independent event
sets and combined to improve accuracy.  To reduce systematic uncertainties, leakage estimates were
calculated on a detector-by-detector basis (index $i$ below) and where possible the relevant event
set was separated into bins (index $j$ below) of energy and approximate event position (see
``face'' bins below).  In every case the leakage can be expressed as: 
\begin{equation} n = \sum_{i,j} N_i s_{ij} \frac{m_{ij}}{M_{ij}},
\label{eq:leak} 
\end{equation} 
where $n$ is the total expected number of \seuh{s} leaking into the WIMP signal region.  The
symbol $N_i$ is the number of NRSS events in the \nr{} band rejected by the timing cut for the
$i\text{th}$ detector.  $M_{ij}$ and $m_{ij}$ are the number of multiples in (or around; see
below) the WIMP signal region failing and passing the timing cut, respectively.  The $s_{ij}$ are
the fractions of $N_i$ in subset $j$, which are calculated using the \se{} multiples in the \nr{}
band for the WIMP-search data.  

The 14 detectors used in this analysis were split into two detector sets according to their
positions in the tower: the 12 interior detectors, and two ``endcaps,'' on the top or bottom of a
stack of six detectors.  Equation~\ref{eq:leak} was applied to each detector set separately
because endcaps require additional systematic uncertainty corrections (discussed below).  The
$M_{ij}$ and $m_{ij}$ are determined independently for three different event sets: 1) NRMS in
WIMP-search data,  2) WBMS in WIMP-search data, and 3) WBMS in $^{133}$Ba calibration data.  For
representative \se{} populations drawn from outside the \nr{} band, the index $j$ specifies one of
six subsets constructed from two detector faces (see \sect~\ref{sec:quality_cuts} on face tagging)
and three energy bins (10--20\,\keVr{}, 20--30\,\keVr{}, and 30--100\,\keVr{}).  In the case where
$M_{ij}$ and $m_{ij}$ were drawn from the NRMS WIMP-search data, the six event subsets were
combined for each detector because of the low statistics, reducing the index set for $j$ to one
element.  Each of the three independent estimations of $m_{ij}$ and $M_{ij}$ are used with
\eq~(\ref{eq:leak}) to obtain three estimators for $n$, which were then statistically combined to
produce the final leakage estimate. 

With these estimators, Monte Carlo (MC) simulations and Bayesian inference were used to estimate
the \se{} background and its uncertainties after unblinding the data~\cite{JeffThesis}. Instead of
simulating the posterior distribution of $n$ (\eq~(\ref{eq:leak})), for each event set, MC
simulations are run for the posteriors of the individual Poisson counts $m_{ij}$ and $M_{ij}$, and
the multinomial fractions $s_{ij}$. Jaynes priors, $p(\rho) \propto \rho^c$ with $c\approx-1$,
were chosen for the Poisson distribution. This prior is generally considered an ``objective
prior,'' having the advantage of ensuring invariance of statistical inference under
transformations of the Poisson mean~\cite{PDGSTATS2012}. A uniform prior was used for the
multinomial distribution. The posterior of $n$ was then calculated using \eq~(\ref{eq:leak}) with
the simulated posteriors of each component.

Systematic uncertainties of different origins were estimated independently, added in quadrature,
and then incorporated into the posterior of the leakage with a profile of the standard normal
distribution.  Systematic uncertainties from two sources were estimated, for all the detector and
data-set combinations, including the choice of the prior exponent $c$ and the difference between
singles and multiples.  

End-cap detectors can have their multiples tagged only on one side.  This biases the counts of
singles upward in these detectors.  Together with the low number of \seuh{s} in the data, it is
challenging to estimate the \se{} background. As a result, a conservative estimate where we allow
the $s_{ij}$ to be biased higher for endcaps--the detectors with the worst leakage--is used.
As a double check, the component leakages for each detector obtained by the Bayesian approach were
compared to those obtained by the frequentist approach~\cite{Ruchlin2012}. There is good agreement
between the two.

The final \se{} leakage estimates incorporating both systematic and statistical uncertainties are
displayed in \tab~\ref{tab:allleakage} for all timing cuts presented in this work.  The analyses
with the best expected sensitivities give leakage estimates of 1.19\,(+0.23\,-0.20) events
(\fivedchi{} for the 10\,\keVr{} threshold analysis) and $>$\,1.48\,(+0.20\,-0.20) events (classic
extended threshold).   The leakage estimate for the extended-threshold analysis is a lower limit
because our standard background sample does not extend all the way down to the signal-region
threshold.  WIMP-search multiple scatters below the \er{} mean -5$\sigma$ line and above the \nr{}
mean +2$\sigma$ line (WBMS) is one of our typical background samples, but because the \er{} mean
-5$\sigma$ and \nr{} band lines cross there is an implicit energy threshold in this background
sample.  The signal region, however, is taken to be the \nr{} band above the 3$\sigma$ charge
threshold and below the \er{} mean -3$\sigma$ line.  This leaves a small region (see
\fig~\ref{fig:c58r_bands}) that is unaccounted for in the background estimates.  The estimate is
nevertheless useful because 1) the implicit background-estimation threshold can be as low as
$\sim$5\,\keVr{}, very near the signal-region threshold in most cases, and 2) the limits are
calculated without background subtraction so a conspicuous rise in the WIMP upper limit would
exist if there were a large leakage present in the unaccounted region. 

\subsection{\label{sec:cosneutron_bknds}Cosmogenic background estimate}
To estimate the cosmogenic neutron background, a combination of simulation and data was
used~\cite{AgneseCDMSCosmoNeut}.  Parent muons that intersect the \cdmstwo{} 5-cm-thick
scintillator panels are vetoed at nearly 100\% efficiency. Our data contains only a small number
of neutrons that accompany identified muons, so a high-statistics simulation was used to estimate
how many veto-coincident neutrons produce \nr{s} in the Ge ZIP detectors, compared with those that
are not accompanied by veto activity.  

Simulation muons generated using the MUSUN~\cite{Kudryavtsev2009} program based on slant-path data
from Soudan2 angular muon flux measurements~\cite{SueThesis} were used as input to a GEANT4 MC
model of the Soudan \cdmstwo{} experimental setup.  These muons have a mean energy of
$\sim$215\,GeV and azimuthal and zenith distributions characteristic of the overburden of the
Soudan site and depth (2090\,m.w.e.).   The simulated muons were generated with the appropriate
angular distributions and spectra on a five-sided parallelepiped (no floor).  They were then
propagated via GEANT4 through 10\,m of rock into the Soudan hall.  All secondaries and the parent
muon(s) were tracked through a complete geometry of the \cdmstwo{} shielding and detector towers.
In the simulations, the CDMS setup is located asymmetrically 2\,m from one wall of the 8\,m
east-west cavern dimension to reproduce albedo effects.  The statistics correspond to 66 live
years.  A multiple is defined in the simulation output as an \nr{} in a Ge detector accompanied by
energy deposition of any sort above 2\,keV in any other ZIP detector.  

In an effort to balance the statistical and systematic uncertainties, three complementary
estimators for the total number of unvetoed single-scatter events were constructed.  For the first
two estimators, the simulation was used to calculate a veto ratio (unvetoed to veto-coincident
\nr{s}) that was then normalized to the veto-coincident data from the WIMP-search measurements to
establish the background estimate.  The second estimator makes use of the higher-statistics
multiple-scatter sample, whereas the first uses only single scatters.  The third estimator uses
only simulated unvetoed single-scatter events, scaled to the correct experimental live time, and
accounts for the detector efficiencies for each timing analysis. 

The veto ratio is \ne{0.008}{0.003} for single scatters and \ne{0.003}{0.001} for multiple-scatter
\nr{} events in the simulation.  One might expect a smaller veto ratio for multiple-scatter events
if they are taken to represent more pervasive showers; excess energy from those showers will often
be detected by the veto system even when the parent muon misses the scintillator panels.  The
weighted average for all events gives a veto ratio of \ne{0.004}{0.001}.

There are 14 Ge \nr{} events in our veto-coincident data, of which three are single scatters.  This is
in agreement with the MC which predicts that $\sim$26\% of the vetoed events should be singles,
and the rest multiples.  Of the vetoed \nr{} events, five pass the \fivedchi{} timing cut and four pass
the classic or neural-network timing cuts.  For the data-driven estimators, we normalize to those
passing the timing cut, rather than introduce additional systematic uncertainty by applying an
average timing-cut efficiency to all 14 detectors in the simulation.  However, the numbers
obtained either way are consistent, with trade-offs between systematic uncertainty and
statistical uncertainty in each case.  We decided to quote the simulation-driven estimate because
it is consistent with the others and offers the best statistical uncertainty.  The systematic
uncertainty is taken as the spread between all the estimators.

Based on the 16 unvetoed \nr{} singles over 66 live years of simulation data, the \fivedchi{}
timing-cut cosmogenic neutron background is \nes{0.021}{0.008}{0.009} events for the 10\,\keVr{}
analysis. The classic and neural-network cuts give \nes{0.019}{0.007}{0.01} events and
\nes{0.018}{0.007}{0.01} events respectively for the 10\,\keVr{} threshold case.  

The estimates were also made for the extended-threshold analyses by using the energy range
2--20\,\keVr{} to approximate the extended-threshold energy ranges.  Based on 12 unvetoed \nr{}
singles in that range over 66 live years the \fivedchi{} extended timing-cut cosmogenic neutron
background is \nes{0.009}{0.004}{0.001} events, the classic is \nes{0.012}{0.005}{0.004} events,
and the neural network is \nes{0.014}{0.006}{0.009} events.

\subsection{\label{sec:radneutron_bknds}Radiogenic background estimate}
The radiogenic neutron background was also estimated using a GEANT4 simulation of the \cdmstwo{}
tower configuration.  Neutrons originating from the Th and U decay chains were simulated for each
material in the setup (lead, polyethylene, copper).  The primary energy spectra of the neutrons
was generated using the SOURCES4C package~\cite{Wilson2002,Wilson2005117,Charlton19981}, which
computes the neutron spectra due to spontaneous fission and $\left(\alpha,n\right)$ reactions of
alphas from the full decay chains within the matrix material.  The neutrons were propagated
through all materials, eventually creating \nr{s} in the Ge and Si detectors.  The single- and
multiple-scatter rates were tabulated for each detector to create high-statistics files labeled by
their contaminant source. The files were then weighted by the contamination level (see below) and
normalized to the amount of material present to determine the final background rate.  The source
with the highest single-scatter contribution is $^{238}$U in the copper cryostat enclosures
(cans), followed closely by U and Th contaminants in the lead shielding.

The contamination levels were determined by a separate $\gamma$ simulation, again using GEANT4 and
the same \cdmstwo{} geometry.  In this simulation, gammas from the $^{232}$Th and $^{238}$U decay
chains, $^{40}$K, and $^{60}$Co were generated from inside the shielding components and tower
structures, and from radon daughters on surfaces.  In order to reproduce the fiducial-volume cut
already applied to the data, the location of the energy deposition and the electric field map in
the detectors were used to deduce the fraction of charge collected on the inner and outer
electrodes. The event was cut if it produced a measurable signal on the outer electrode.  A charge
threshold was implemented in the simulation data by using the experimental inner-electrode charge
thresholds (in \keVee) and applying them to the inner-electrode energy derived from the above
procedure.  

\begin{figure}[!htb]
   \includegraphics[width=\columnwidth]{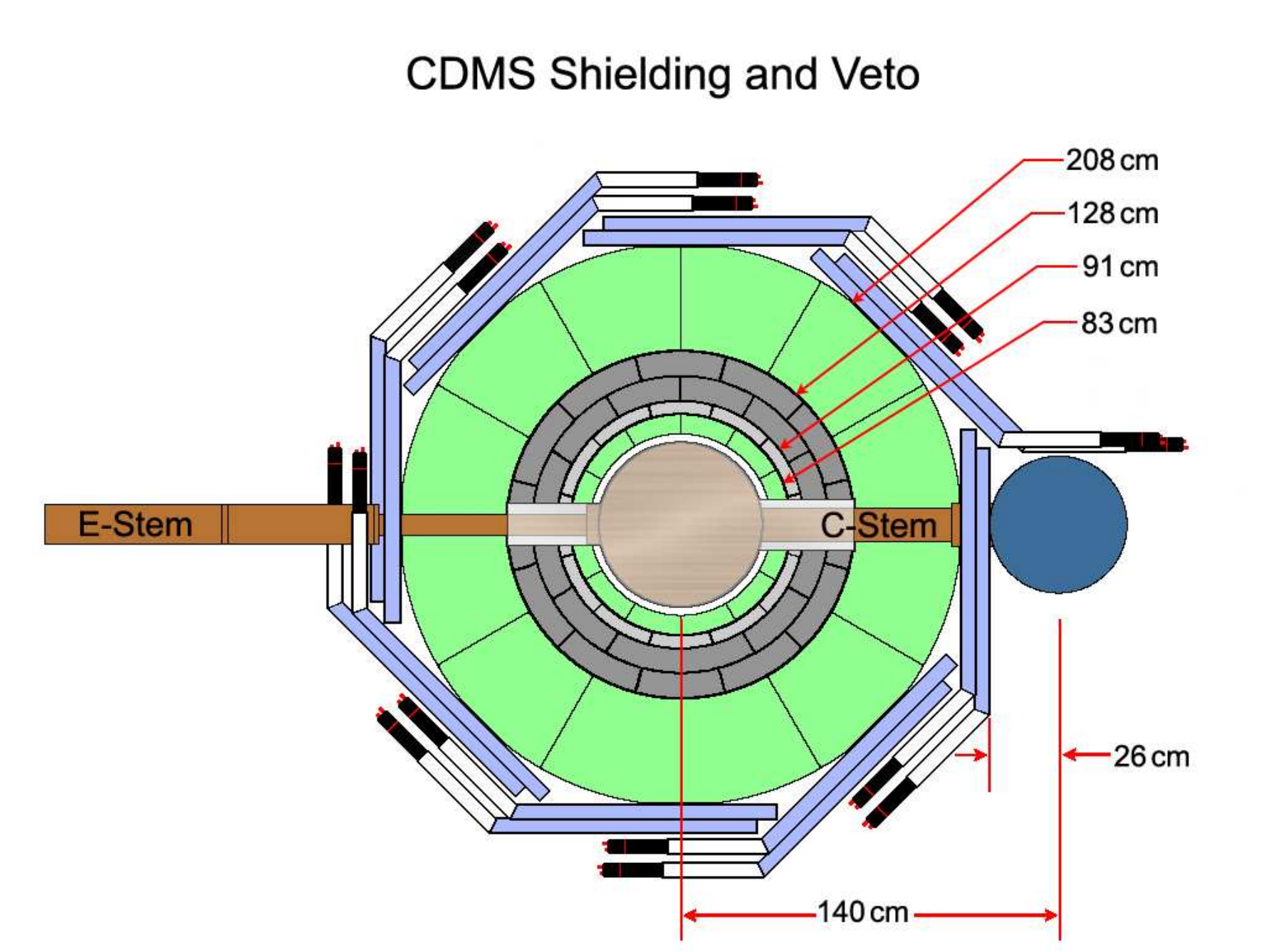}
   \caption{\label{fig:vetotop}(Color online)  Schematic of the \cdmstwo{} shielding
   configuration.  The outermost vacuum can surface is shown near the center and radially outward
   the layers are:  inner polyethylene (green), inner lead (light grey), outer lead (darker grey),
   outer polyethylene (green), and muon-veto scintillator counters (blue) with attached light
   guides and phototubes (white and black). Cooling of the detectors is achieved through the
   ``cold stem'' from a dilution refrigerator (dark blue), while cabling passes through the
   ``electronics stem'' on the other side of the setup.
   }
\end{figure}

\begin{figure*}[!htb]
   \includegraphics[width=\textwidth]{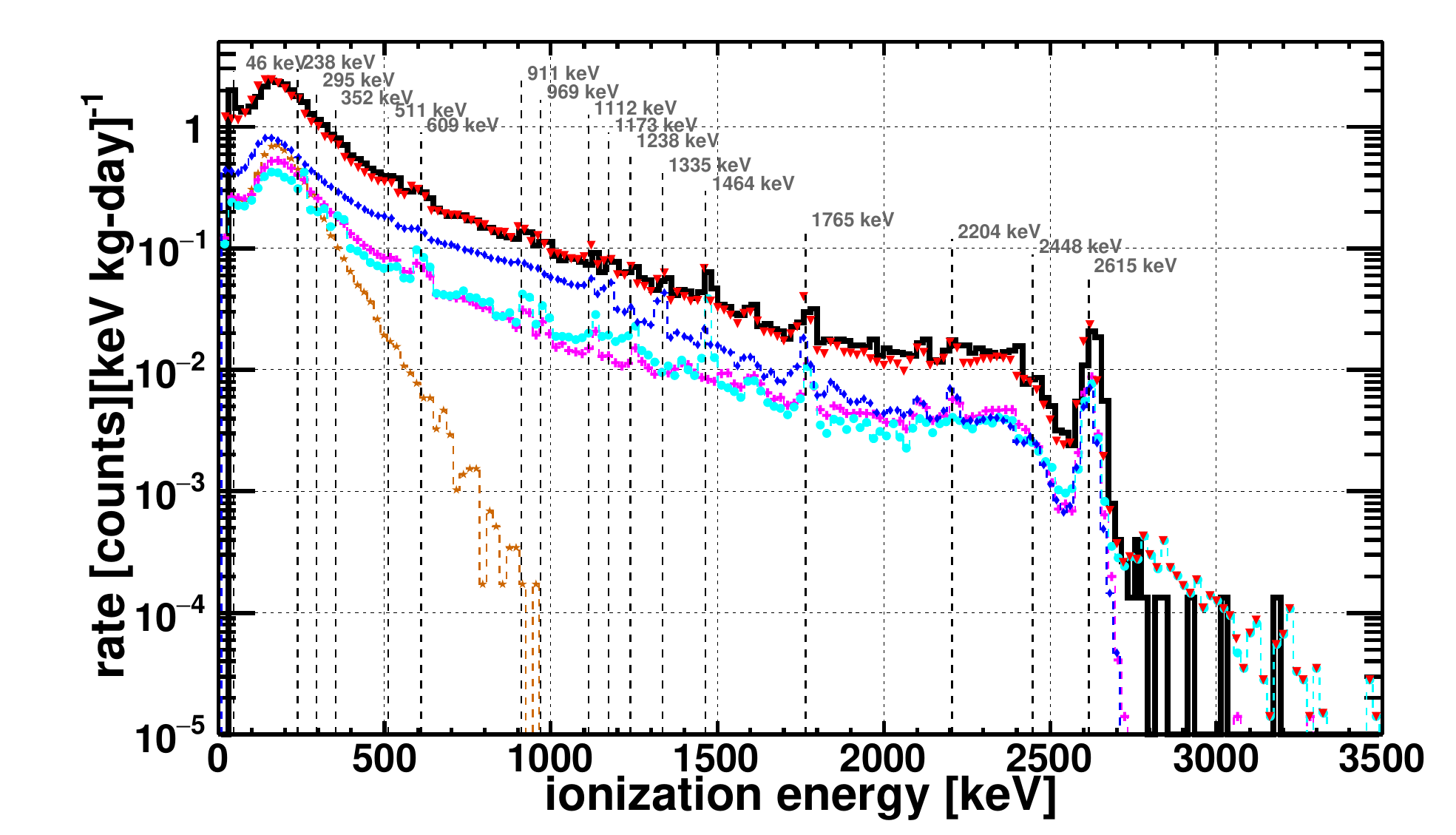}
   \caption{\label{fig:gamma_spectrum}(Color online) Measured event-summed energy across all
   detectors (black solid line) compared to simulations of contaminants from common locations:
   lead shielding (blue diamonds), inner polyethylene shielding (magenta crosses), copper cans and
   tower components (cyan filled circles), and outermost can surfaces (brown stars). The sum of
   all simulated components (red inverted triangles) is also plotted and is in good agreement with
   the measured data.
   }
\end{figure*}

To obtain the contamination estimate a $\chi^2$ minimization was performed to fit the 43 sources
considered (see \fig~\ref{fig:vetotop} for a schematic of the setup and \tab~\ref{tab:mc_sources}
for a breakdown of source locations) to the \er{} data above 15 \keVee, avoiding the
$\sim$10\,\keVee Ge activation peaks.  Some representative values for this contamination are
4\,mBq/kg for both $^{232}$Th and $^{238}$U in the inner polyethylene shield and 6--7\,mBq/kg for
$^{232}$Th and $^{238}$U in the tower 5 copper, the tower with the highest contamination.  The
comparison between the \er{} spectrum and the sum of MC results using the best-fit contamination
levels is shown in \fig~\ref{fig:gamma_spectrum}.

\begin{table*}[!htb]
\begin{tabular}{ c | c | c | c | c | c | c | c }
\hline
\hline
Contamination & Outer Lead & Inner Lead & Inner Polyethylene &  Cans & Innermost Can Surface & Outermost Can Surface & Towers\\ \hline
$^{60}$Co & \checkmark & - & - & \checkmark & \checkmark & \checkmark & \checkmark\\ 
$^{40}$K & \checkmark & \checkmark & \checkmark & \checkmark & \checkmark & \checkmark & \checkmark\\ 
$^{232}$Th & \checkmark & \checkmark & \checkmark &  \checkmark & \checkmark & \checkmark & \checkmark\\ 
$^{238}$U & \checkmark & \checkmark & \checkmark &  \checkmark & \checkmark & \checkmark & \checkmark\\ 
$^{222}$Rn & - & - & - & - & - & \checkmark & -\\ 
\hline 
\hline 
\end{tabular}
   \caption{\label{tab:mc_sources}Sources used for the radioactive contamination fitting
   procedure.  The \checkmark indicates that the respective source was used in the fitting
   procedure for the location given.  The cans are the nested copper cold stages of the cryostat,
   where surface sources were simulated on both the innermost and outermost surfaces.  There are
   five separate sources for the copper components of the five detector towers.  No simulated
   sources were placed in the outer polyethylene or the veto panels because they are behind too
   much shielding to yield a measurable amount of background.  
   }
\end{table*}

Because of nearly degenerate fits to the spectrum, there is uncertainty in the final contamination
values.  The degree of uncertainty was estimated by bracketing the largest changes that produce
acceptable fits in the gamma MC and translating these into the resulting changes in the neutron
rate.

In the 14 Ge detectors used for the WIMP-search analysis, the expected raw radiogenic neutron
single-scatter background rate was found to be (\ne{1.15}{0.14})$\times10^{-4}$ events\perkgd{} in
the 10--100\,\keVr{} energy range.  The statistical uncertainty is negligible because of the large
number of events simulated while the systematic uncertainty listed is due to the uncertainty in
the contamination values and locations.  This singles spectrum was then convolved with the final
neutron cut efficiencies from the three timing analyses (see \fig~\ref{fig:timing_effs}).  The
radiogenic spectrum-averaged efficiencies are: 35.3\% for the classic, 34.9\% for the neural net,
and 40.4\% for the \fivedchi{} analysis.  For the raw exposure of 612.2\,\kgd, the final
radiogenic neutron event backgrounds for the three analyses are: \nes{0.025}{0.0001}{0.003} events
for the classic and neural network, and \nes{0.028}{0.0001}{0.004} events for the \fivedchi{}
analysis.

For the extended analyses the estimates were made with the approximate energy range
2--20\,\keVr{}.  The estimates were: \nes{0.0131}{0.0001}{0.0019} for the classic,
\nes{0.0148}{0.0001}{0.0021} for the neural network, and \nes{0.0105}{0.0001}{0.0015} for the
\fivedchi{}.

\subsection{\label{sec:pb_bknds}$^{206}$Pb background estimate}
$^{210}$Po decays via the $\alpha$-decay process. This leaves an $\alpha$ particle of 5.4\,MeV and
a recoiling ($\lesssim$100\,keV) $^{206}$Pb nucleus in the final state.  $^{210}$Po is in the
$^{238}$U decay chain and is part of possible post-$^{222}$Rn plate-out contamination. It is also
the first post-$^{222}$Rn $\alpha$-decayer after the long-lived $^{210}$Pb. These facts make this
decay likely to be an important long-lived contamination, and because a recoiling $^{206}$Pb
nucleus could have produced an energy deposition below $\sim$100\,\keVr{} with an ionization yield
consistent with being an \nr{}, it makes sense to evaluate this background very seriously.

Because of the \cdmstwo{} detector geometry (see \fig~\ref{fig:stack}) much of the surrounding
surface of an interior detector is comprised of another detector.  $^{210}$Po decay events in
which the decay occurs in one detector and the $^{206}$Pb recoil is registered in an adjacent
detector cannot contaminate the signal region of the latter detector because of the clear 5.4\,MeV
$\alpha$ deposition in the former.  Therefore the most important component of the background comes
from surfaces that are uninstrumented or those that are adjacent to uninstrumented surfaces.
Another important point is that since the alphas from this decay give such a clear signature, and
the decay is a two-body decay, an obvious way to estimate the number of unaccompanied (single)
$^{206}$Pb events is to estimate the number of unaccompanied $\alpha$ events from the decay. The
angular distribution is isotropic and any surface that can observe a single $\alpha$ is
approximately equally likely to observe a single $^{206}$Pb recoil.  This estimation of the number
of single $^{206}$Pb events was carried out, and using rough estimations for the passage fractions
of these events when subjected to the other analysis cuts, we expect approximately 0.187
signal-region $^{206}$Pb events over the whole exposure.  

Since the $^{206}$Pb recoil estimates were inferred from $\alpha$ counting, and the $^{206}$Pb
recoils can come at different energies depending on how deeply the $^{210}$Po parent is embedded
into the originating surface, we do not have a good specification of the energy distribution of
such events.  Therefore, to be conservative we can use the same estimate for the 10\,\keVr{} and
extended-threshold analyses.  We expect \nes{0.187}{0.018}{0.187} events over the whole exposure;
and have assigned a 100\% systematic uncertainty to account for the roughness of this estimate.
This background estimate is clearly subdominant with respect to the \se{} background estimates,
but is larger than the cosmogenic or radiogenic neutron background estimates~\cite{TommyThesis}.

\section{\label{sec:results}Results}
The background estimates for the primary 10\,\keVr{} and extended-threshold analyses are
summarized in \tab~\ref{tab:summarized_bknds}.  While the background estimates can be used to
interpret the overall results of the experiments they do not directly modify the limit curves (see
below).
\begin{table}[!hbt]
\begin{tabular}{ c  c  c }
\hline
\hline
Background	& \fivedchi{} (10\,\keVr{}) & Classic (extended)\\ \hline
Leakage		& \ne{1.19}{0.22}  		&  $>$\ne{1.48}{0.20}  \\
Cosmogenic	& \ne{0.021}{0.012}     &  \ne{0.012}{0.006}  \\ 
Radiogenic	& \ne{0.028}{0.004}   		&  \ne{0.013}{0.002}   	 \\
$^{206}$Pb	& \ne{0.19}{0.19}		&  \ne{0.19}{0.19}  \\ \hline
Total		& \ne{1.43}{0.30}  	   	&  $>$\ne{1.69}{0.28} \\
\hline
\hline
\end{tabular}
   \caption{\label{tab:summarized_bknds} A summary of the background estimates from the primary
   analysis methods. Statistical and systematic uncertainties have been added in quadrature and to
   compute the total backgrounds all component uncertainties were also added in quadrature. 
   }
\end{table}

Results of direct WIMP-search experiments are usually summarized as upper limits or signal
contours in the WIMP-nucleon cross section versus WIMP-mass plane. Yellin's optimum interval
method~\cite{Yellin2002} allows derivation of an upper limit on a signal rate in cases with
unknown background. While the backgrounds in our signal region are not completely unknown, this is
a conservative approach to setting upper limits on a possible signal. This presentation also
requires assumptions about the WIMP distribution in the galactic halo, the type of interaction
between WIMPs and nucleons, and the nuclear form factor for the interaction.  The velocity
distribution was assumed to be Maxwellian and was parametrized by the rotational velocity at
infinite radius and corrected for the finite galactic escape velocity~\cite{Lewin1996}, which is
taken to be 544~km/s~\cite{Smith2007}.  A WIMP mass density of 0.3~GeV/$c^2$/cm$^3$ was used for
historical reasons, making the computed limits comparable to similar
publications~\cite{Lewin1996}.  Some recent astrophysical measurements indicate different
values~\cite{Lavalle2015};  a correction for such deviations is a simple multiplicative factor and
can be easily applied by the reader~\cite{Bernabei1996757}.  A most probable WIMP velocity of
220~km/s was used along with a mean circular velocity of the Earth with respect to the Galactic
center of 232~km/s.  The WIMP interactions were assumed to be spin independent and the Helm form
factor was used~\cite{Lewin1996} for a natural Ge isotopic distribution.  

\begin{figure}[h!tb]
\includegraphics[width=\columnwidth]{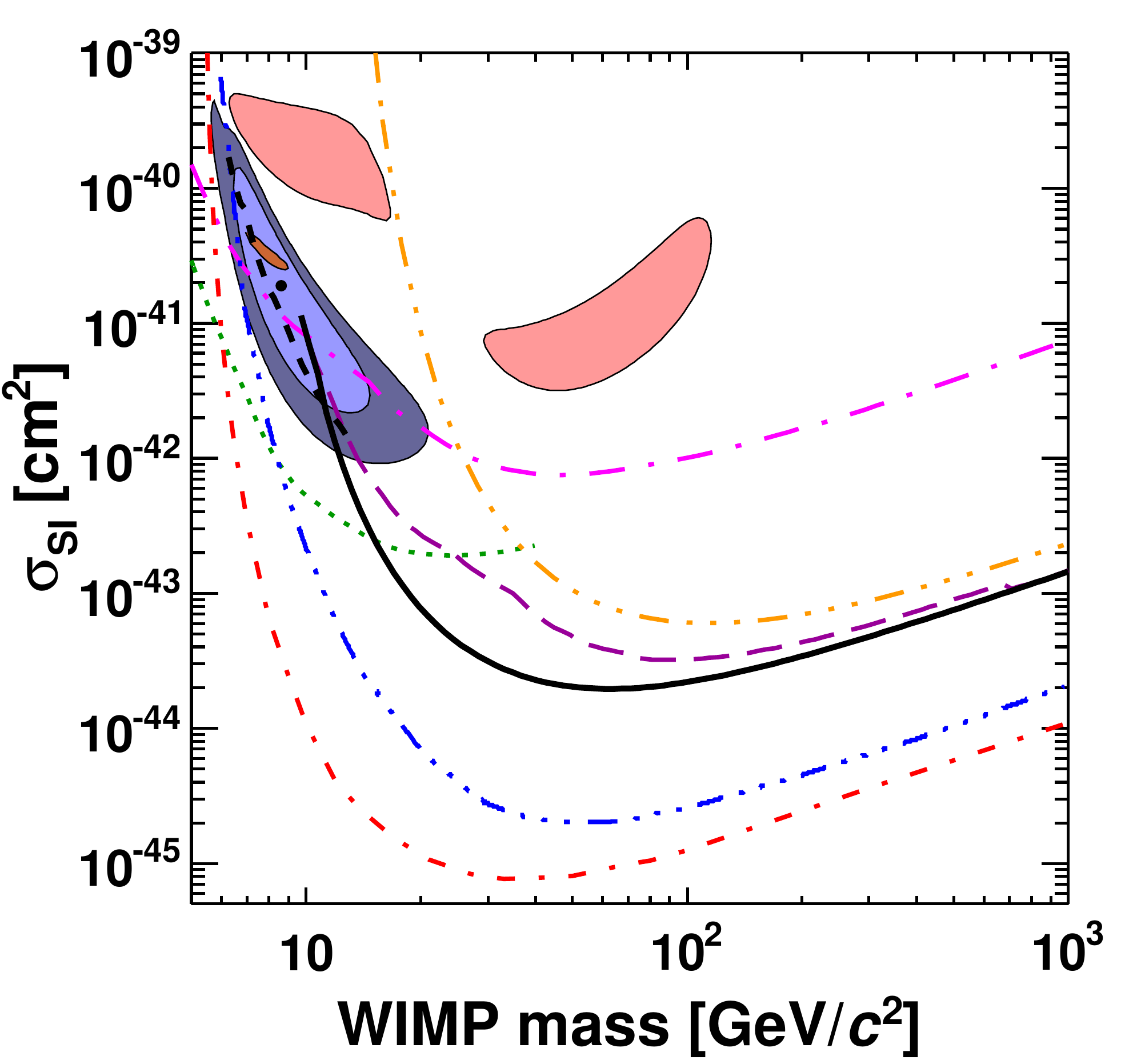}
   \caption{\label{fig:combined_c58r_limits} This figure compares the main results from this
   analysis (the current \fivedchi{} analysis combined with prior \cdmstwo{} exposures, black
   solid and the classic extended-threshold analysis, black dashed) with previously published
   results (all limits are at 90\% C.L.): Darkside-50~\cite{Agnes2015456} (orange
   triple-dot-dashed) XENON100~\cite{Aprile2012} (blue triple-dot-dashed); LUX~\cite{Akerib2013}
   (red dot-dashed); SuperCDMS low-threshold~\cite{Agnese2013LT} (green dotted);
   \cdmstwo{}/EDELWEISS combined~\cite{CDMS_EDEL_2011} (purple long dashed);
   CRESST-II~\cite{CRESSTII_2015} (magenta long-dot-dashed); \cdmstwo{} Si~\cite{Agnese2013} (90\%
   and 68\% C.L. contours, blue, with the best-fit point marked with a black dot);
   DAMA/LIBRA~\cite{Bernabei2010,Savage2009} (3$\sigma$ region, light red) and
   CoGeNT~\cite{Aalseth2013} (90\% C.L., brown).
   }
\end{figure}

The comparison of the present 10~\keVr{} \fivedchi{} and classic extended results with other
published limits and signal contours is shown in \fig~\ref{fig:combined_c58r_limits}.  In the
figure, our 10~\keVr{} \fivedchi{} limit is combined with the \cdmstwo{} five-tower exposure
acquired before July 2007, resulting in a limit that summarizes the full (and final) \cdmstwo{}
high-threshold sensitivity.   The \cdmstwo{}/EDELWEISS combined limit~\cite{CDMS_EDEL_2011} is
also shown for comparison.  Above $\sim$100~\gev{} WIMP mass the combined limit is comparable to
our \cdmstwo{} combined result owing to the good efficiency-averaged exposures of both of the
experiments in the relevant energy ranges. 

\subsection{\label{sec:indiv_limits}Limit cross-checks}
To gain insight into the effect of timing-cut figures of merit (efficiency, SAE, leakage) on
WIMP-search results, we have constructed final limits for all of the timing-cut constructions in
this work.  Although the SAE and expected leakage were exclusively used to choose the primary
timing cuts, comparing all of the limits in this way gives cross-checks on how other parameters
(efficiency, signal-region events) affect the reach of the experiment. 

\begin{figure}[!htb]
  \includegraphics[width=\columnwidth]{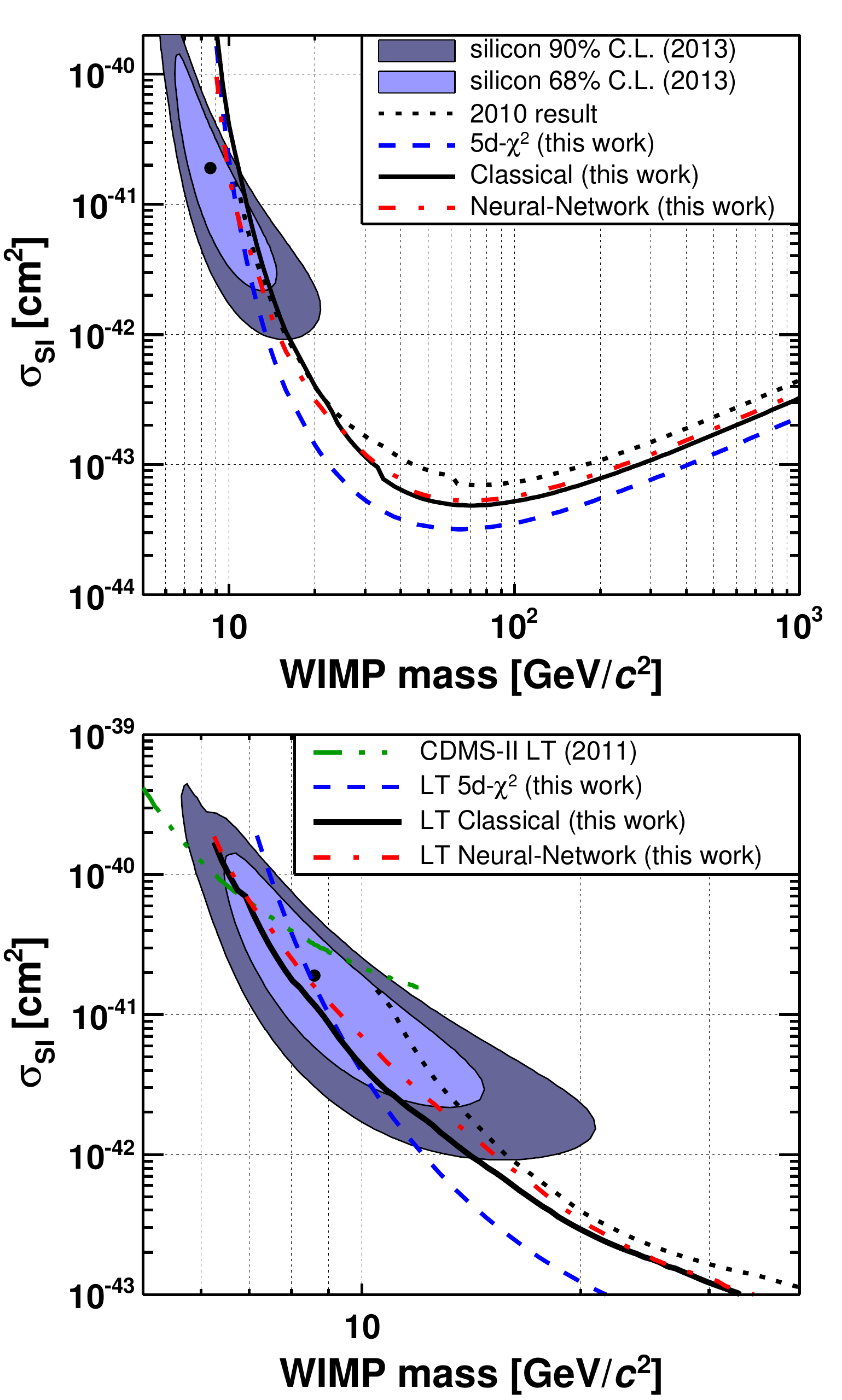}
  \caption{\label{fig:c58r_limits}(Color online)  Experimental upper limits (90\% confidence
  level) derived from each of the analyses presented in this work compared with the originally
  published~\cite{Ahmed2010a} (black dotted) limits.  The \cdmstwo{} Si contour is shown with the
  best-fit point marked with a black dot (WIMP mass of 8.6~\gev{} and WIMP-nucleon cross section
  of 1.9$\times$10$^{-41}$~cm$^2$)~\cite{Agnese2013}.  (Top panel) The 10~\keVr{} threshold
  analyses.  The \fivedchi{} limit (blue dashed) is the ``primary'' high-threshold result to be
  quoted from this work.  The neural-network and classic limits are shown as red dot-dashed and
  black solid lines respectively.  (Bottom panel) The extended-threshold limits, focused on the
  lower WIMP-mass region.  The same color code applies except that all of the analyses from this
  work correspond to the extended-threshold versions.  The classic limit (black solid) is the
  ``primary'' extended-threshold result to be quoted from this work.  The extended \fivedchi{}
  limit shown corresponds to the timing-cut optimization assuming a 60\,\gev{} WIMP mass.  For
  comparison, the previous \cdmstwo{} low-threshold limit is shown~\cite{Ahmed2011} (green
  triple-dot-dashed).
  }
\end{figure}

The top part of \fig~\ref{fig:c58r_limits} shows the limits derived using the optimum interval
method for the different timing cuts with 10~\keVr{} thresholds.  Relative to the original
publication, the ionization-based fiducial-volume and phonon-timing cuts have improved
efficiencies in the analysis reported here, leading to the improved exposure and more stringent
limits.  In terms of the overall spectrum-averaged detection efficiency for a 60~\gev{} WIMP, the
classic timing-cut strategy (see \sect~\ref{sec:classical_timing}) shows a 12\% improvement over
the previously published version, about half of which can be attributed to a reoptimization of the
ionization-based fiducial volume following the data reprocessing.  A similar improvement was seen
for the neural-network timing-cut analysis (described in \sect~\ref{sec:nnet_timing}).  The
largest improvement of 29\% in overall SAE efficiency was achieved for the \fivedchi{} analysis
(described in \sect~\ref{sec:5dchi_timing}), owing primarily to an increased timing-cut efficiency
in the 15--90~\keVr{} energy range (see \sect~\ref{sec:final_timing_efficiency}).

For the 10~\keVr{}-threshold analyses, the \fivedchi{} sets the most stringent limit at a 60~\gev{}
WIMP mass, while the neural-network timing cut results in stronger limits at and below 10~\gev{},
an important region for further study~\cite{Agnese2013}.  The \fivedchi{} set weaker limits for
low-mass WIMPs because the cut was set to maximize sensitivity to a WIMP with mass 60~\gev{}.
This combined with the fact that the \fivedchi{} method could set a very tight cut at low energies
(it used an independent 10--20~\keVr{} bin whereas the other analyses were less granular) produced
a poorer WIMP efficiency toward low recoil energies despite the lower expected background leakage
at those energies, but excellent sensitivity for high WIMP masses.

The lower part of \fig~\ref{fig:c58r_limits} shows the extended limits in the low-WIMP-mass
region.  Each extended analysis constrains the 8--10~\gev{} mass region more strongly than the
higher-threshold analyses, and the classic timing cut produces the strongest limit near the
silicon-detector analysis best-fit point of $M_{\mathrm{W}}=$\,8.6~\gev{} and
$\sigma_{SI}=$\,1.9$\times$10$^{-41}$~cm$^2$~\cite{Agnese2013}.  The extended limits are also
compared to the previous low-threshold \cdmstwo{} results, which did not use a timing
cut~\cite{Ahmed2011}.  That analysis has a larger exposure toward lower recoil energies which
accounts for the stronger limit set below a $\sim$7~\gev{} WIMP mass.  The classic analysis
presented here has a stronger limit by a factor of approximately 2.7 at a WIMP mass of
$\sim$8.6~\gev{}. 

\section{\label{sec:conclusions}Conclusion}
The reprocessed data did not produce significant changes in the number of signal-region events,
indicating that uncertainties applied in the original processing of the \cdmstwo{} data
set~\cite{Ahmed2010a} were robust. All three sets of higher-threshold timing cuts produced similar
limits, with small differences consistent with their corresponding exposure-optimization
procedures.  For example, the \fivedchi{} analysis has a high efficiency at moderate recoil
energies (30--60~\keVr{}), but has a stringent timing cut at lower energies. It is well suited to
provide the strongest limits at high WIMP mass ($>$60~\gev), but will produce fewer low-energy
signal-region events.  On the other hand, the classic analysis at 10~\keVr{} threshold shows a
slight weakening of the 90\% C.L. limit for WIMP masses below about 18~\gev{}, where sharp increases
in the limit curves indicate systematics near threshold.  The neural-network timing cut has been
identified as a robust method with the highest signal efficiency at low energies (see
\fig~\ref{fig:timing_effs}) and good sensitivity at lower WIMP mass ($\leq$~9~\gev).  The classic
extended-threshold limit rules out about half of the silicon 68\% C.L. region obtained in a
previous \cdmstwo{} publication~\cite{Agnese2013}.  This indicates that the low-threshold results
from the Ge detectors are marginally compatible with the Si-detector measurements taken during the
same data period, under standard assumptions.  Comparisons of such direct detection results on
different nuclei will be a powerful tool for understanding WIMP dark matter both in terms of the
fundamental WIMP interactions~\cite{Haxton2014,Schneck:2015eqa} and possible backgrounds.

\begin{acknowledgements}
The CDMS Collaboration gratefully acknowledges the contributions of numerous engineers and
technicians; we would like to especially thank Dennis Seitz, Jim Beaty, Bruce Hines, Larry Novak,
Richard Schmitt and Astrid Tomada.  In addition, we gratefully acknowledge assistance from the
staff of the Soudan Underground Laboratory and the Minnesota Department of Natural Resources.
This work is supported in part by the National Science Foundation, the US Department of Energy,
NSERC Canada, and by MultiDark (Spanish MINECO). Fermilab is operated by the Fermi Research
Alliance, LLC under Contract No. De-AC02-07CH11359. SLAC is operated under Contract No.
DE-AC02-76SF00515 with the United States Department of Energy. 
\end{acknowledgements}

\clearpage
\bibliography{c58R_Ge_allrefs_short}
\bibliographystyle{apsrev4-1}

\end{document}